\documentclass[12pt,preprint]{aastex}
\usepackage{emulateapj5,onecolfloat}
\usepackage{epsfig}
\usepackage{times}

\def\lgl{\langle}
\def\rgl{\rangle}

\def\ltsima{$\; \buildrel < \over \sim \;$} 
\def\simlt{\lower.5ex\hbox{\ltsima}}
\def\gtsima{$\; \buildrel > \over \sim \;$} 
\def\simgt{\lower.5ex\hbox{\gtsima}}
\def\simless{\mathbin{\lower 3pt\hbox
   {$\rlap{\raise 5pt\hbox{$\char'074$}}\mathchar"7218$}}}   
\def\simgreat{\mathbin{\lower 3pt\hbox  
   {$\rlap{\raise 5pt\hbox{$\char'076$}}\mathchar"7218$}}}   

\shorttitle{A weak lensing analysis of the Abell 901/902 supercluster}
\shortauthors{Gray et al.}

\begin{document}

\submitted{Accepted for publication in the Astrophysical Journal, November 10, 2001} 

\twocolumn[ 

\title{Probing the distribution of dark matter in the \\Abell 901/902 supercluster with
weak lensing}
\author{M.E. Gray}
\affil{Institute of Astronomy, Madingley Road, Cambridge CB3 0HA,
United Kingdom}
\email{meg@ast.cam.ac.uk}

\author{A.N. Taylor}
\affil{Institute for Astronomy,  Blackford Hill, Edinburgh EH9 3HJ,
United Kingdom}
\email{ant@roe.ac.uk}

\author{K. Meisenheimer}
\affil{Max-Planck-Institut f\"{u}r Astronomie, K\"{o}nigstuhl 17,
D-69117, Heidelberg, Germany}
\email{meise@mpia-hd.mpg.de}

\author{S. Dye}
\affil{Astrophysics Group, Blackett Lab, Imperial College, Prince
Consort Road, London SW7 2BW, United Kingdom}
\email{s.dye01@ic.ac.uk}

 \author{C. Wolf, \& E. Thommes}
\affil{Max-Planck-Institut f\"{u}r Astronomie, K\"{o}nigstuhl 17,
D-69117, Heidelberg, Germany}
\email{cwolf@mpia-hd.mpg.de}
\altaffiltext{1}{Institute for Astronomy,  Blackford Hill, Edinburgh
EH9 3HJ, United Kingdom}

\begin{abstract}
We present a weak shear analysis of the Abell 901/902 supercluster,
composed of three rich clusters at $z=0.16$.  Using a deep $R$-band
image from the $0.5^\circ\times 0.5^\circ$ MPG/ESO Wide Field Imager
together with supplementary $B$-band observations, we build up a
comprehensive picture of the light and mass distributions in this
region.  We find that, on average, the light from the early-type
galaxies traces the dark matter fairly well, although one cluster is a
notable exception to this rule.  The clusters themselves exhibit a
range of mass-to-light ($M/L$) ratios, X-ray properties, and galaxy
populations.  We attempt to model the relation between the total mass
and the light from the early-type galaxies with a simple
scale-independent linear biasing model.  We find $M/L_B=130h$ for the
early type galaxies with zero stochasticity, which, if taken at face
value, would imply $\Omega_m < 0.1$.  However, this linear relation
breaks down on small scales and on scales equivalent to the average
cluster separation ($\sim$1 Mpc), demonstrating that a single $M/L$
ratio is not adequate to fully describe the mass-light relation in the
supercluster.  Rather, the scatter in $M/L$ ratios observed for the
clusters supports a model incorporating non-linear biasing or
stochastic processes.  Finally, there is a clear detection of
filamentary structure connecting two of the clusters, seen in both the
galaxy and dark matter distributions, and we discuss the effects of
cluster-cluster and cluster-filament interactions as a means to
reconcile the disparate descriptions of the supercluster.
\end{abstract}


] 

\section{Introduction}

Gravitational lensing is a powerful tool that allows us to directly
map dark matter and determine the relation between the observed
distribution of light and the underlying mass distribution. In recent
years, the development of sophisticated lensing techniques coupled
with a new generation of wide-field imaging instruments has opened
doors to lensing studies on unprecedentedly large scales.  Weak
lensing is now a well-established method for determining the mass
distribution of rich clusters of galaxies \citep{bs01,mellier99}.
With the increased confidence in lensing techniques, recent studies
have turned from clusters to blank fields
\citep{bacon00,kaiser00,vanwaerbeke00,wittman00,maoli01} to measure
the statistical `cosmic shear' due to lensing by large-scale
structures.

Superclusters, comprising $\sim$2-10 clusters of galaxies, are the
largest known systems of galaxies in the Universe \citep{vogeley94}
and represent a stepping-stone between rich clusters and large-scale
structure.  Clusters of clusters and associated filamentary structures
in the galaxy distribution have been observed at low redshift through
redshift surveys of the Corona Borealis, Shapley, and Perseus-Pisces
superclusters \citep[e.g.][]{small97,quintana95,postman88}.  Recently,
supercluster studies have been extended to intermediate redshifts with
the weak lensing study of MS0302+16 at $z=0.4$ by \cite{kaiser98}, and
pushed to even higher redshifts with the optical detection of a $z
\sim 0.91$ supercluster by \cite{lubin00}.

Superclusters are invaluable testing grounds for theories of
cosmology, the growth of structure, galaxy formation, and the nature
of dark matter.  The presence or absence of filamentary structure can
be used to probe theories of structure formation \citep[e.g.][]{cen95}.
The degree of substructure is reflective of the universal matter
density parameter, $\Omega_m$, \citep{richstone92}, and
simulations find that the mass fraction in filaments is expected to be
slightly larger in a low-density universe \citep{colberg99}.
Furthermore, supercluster mass-to-light ratios ($M/L$) can be also be
used to determine $\Omega_m$ \citep{kaiser98}.  This latter
calculation rests on the assumption that the supercluster $M/L$ ratio
is representative of the Universe as a whole, i.e. that the $M/L$
ratio flattens to a constant on supercluster scales \citep{bahcall00}.

In this paper we present a weak lensing analysis of the Abell
901(a,b)/Abell 902 supercluster (referred to hereafter as A901/902)
using the $0.5^\circ\times0.5^\circ$ Wide-Field Imager
\citep[WFI;][]{WFI1,WFI2} on the MPG/ESO 2.2m telescope.  Our aim is
to trace the dark matter distribution in the supercluster field and to
investigate whether it is concentrated in cluster cores or distributed
through filamentary structures.  In the `Cosmic Web' theory of
structure formation \citep{bond96}, filaments emerge from the
primordial density field after cluster formation.  These filaments are
thought to harbor a large fraction of present day baryons in the form
of hot gas, prompting calls for X-ray searches \citep{pierre00}.
However, attempts to detect filamentary X-ray gas have to date been
unsuccessful, yielding only upper limits \citep[e.g.][]{briel95}.  As
X-ray emissivity scales as the square of the density while
gravitational shear scales linearly, a weak lensing-based approach may
prove more successful in uncovering such structures.  \cite{kaiser98}
report tantalising but tentative evidence for a lensing detection of a
mass bridge within MS0302+16.  Here we will use weak shear
reconstruction algorithms to create two-dimensional maps of the mass
distribution and search for similar filamentary structures within the
A901/902 supercluster region.

In addition, we also wish to determine how well the optical light
traces the underlying dark matter, and how the cluster masses
determined by weak lensing compare to those predicted from the results
of a previous X-ray study of the region \citep{schindler00}.  Finally,
we wish to examine the ratio of the total mass of the system to the
$B$-band light of the early-type supercluster galaxies, and to discuss
the resulting implications for $\Omega_m$.  Measurements of the
mass-to-(total)-light ratio of clusters of galaxies typically yield
values in the range $M/L\sim 200-300h$ \citep{carlberg96,mellier99},
which, if taken as the universal $M/L$, imply $\Omega_m=0.2$.
However, if the $M/L$ ratio continues to increase with scale to the
regime of superclusters or beyond, $\Omega_m$ could be considerably
larger.  This would require the galaxies to be significantly biased
with respect to the mass, so that the efficiency of galaxy formation
is greatly enhanced in the densest regions.

\cite*{bahcall95} explore this hypothesis, and compile
results from the literature to trace an increasing $M/L$ ratio as they
proceed in scale from galaxies to groups and then to clusters.  Their
best-fit values for the $M/L$ of individual galaxies result in a $M/L$
ratio for ellipticals that is $\sim 4$ times that for spirals,
suggesting that on average, ellipticals contain more mass than spirals
for the same luminosity and radius.  They conclude that the total mass
of groups, clusters and superclusters can be accounted for by the mass
of the dark halos of their member galaxies (which may have been
stripped off in a dense environment but still remain within the
system) plus the mass of the hot inter-cluster gas, and that there is
no great repository of dark matter hidden on larger scales.

Indeed, on the scale of superclusters, measurements of galaxy velocity
dispersions indicate that the universal $M/L$ function appears to
flatten to $\sim 300h$ (assuming that the systems are bound but not
necessarily in an equilibrium state).  \cite{bahcall95}
find little evidence that the $M/L$ ratio continues to increase with
scale, implying $\Omega\sim 0.2$.  The issue of a low supercluster
$M/L$ ratio was revisited by \cite{kaiser98} in a weak-shear study
of a supercluster of three X-ray luminous clusters at $z=0.4$. This was
the first direct mapping of dark matter on these scales,
independent of biasing assumptions.  \citeauthor{kaiser98} present the
remarkable result that the light from the color-selected
\textit{early-type galaxies alone} is sufficient to trace the mass as
revealed by weak lensing, with the mass no more extended than the
early-type galaxies and with late-type galaxies having negligible
$M/L_B$.  Our weak lensing study affords us the opportunity to test
this hypothesis on another supercluster system at lower redshift, and
to determine if a single, scale-independent $M/L$ ratio can describe
the relation between mass and light in the entire system.  We will
discuss the resulting implications of such a conclusion for values of
$\Omega_m$, and explore other possible biasing relations.

\subsection{The A901/902 supercluster}

The A901/902 supercluster is composed of three clusters of galaxies,
all at $z=0.16$ and lying within $30\arcmin\times30\arcmin$ on the
sky.  Abell 901 is listed as a double cluster (A901a and A901b) with
irregular morphology in the X-ray Brightest Abell-type Cluster Sample
\citep[XBACS;][]{ebeling96} of the ROSAT All-Sky Survey, having
$L_X$(0.1-2.4 keV)$=6.01$ and $3.49\times 10^{44}$ erg s$^{-1}$.

Pointed ROSAT HRI observations by \cite{schindler00} reveal that the
emission from the `brighter subcluster' of Abell 901 (labelled Abell
901a in the \citeauthor{ebeling96} nomenclature used in this paper) suffers
from confusion with several X-ray point sources in the vicinity.  She
concludes that the X-ray emission from Abell 901a is in fact
point-like and suggests emission from an active nucleus as a likely
candidate for the source.  Abell 901b, on the other hand, is shown to
exhibit very regular and compact cluster emission, with a revised
$L_X$(0.1-2.4 keV)$=3.6\pm 0.1\times 10^{44}$ erg s$^{-1}$, possibly
containing a cooling flow.  No X-ray flux is detected at the optical
center of Abell 902 \citep*[as determined by][]{aco89}, although
there is point-like emission 2 arcmin to the west of this position.

Both the clustering of number counts  \citep{aco89} and the X-ray
emission \citep{ebeling96,schindler00} in this field indicate
that this is an overdense region.  How exactly mass is distributed
through the field and which of the objects discussed here deserves to
be labelled a `cluster' is not so clear, however.  We therefore turn
to gravitational lensing to trace the underlying mass distribution of
this system.

The structure of the paper is as follows: in $\S\ref{sec:obs}$ we
outline the observations of the supercluster field and discuss the
data reduction and astrometric issues.  In $\S\ref{sec:psf}$ we perform
the corrections for the point-spread function necessary to accurately
measure the shapes of the background galaxies.  In
$\S\ref{sec:clusters}$ we discuss the photometric properties of the
clusters, and in $\S\ref{sec:lensing}$ we present the results of the
weak lensing analysis.  $\S\ref{sec:xcorr}$ contains a statistical
cross-correlation of the light and mass distributions.  Finally, in
$\S\ref{sec:discussion}$ we summarize the results and present our
conclusions.

\section{Observations and Data Reduction}\label{sec:obs}

\subsection{Observing strategy}

The observations presented in this paper were undertaken as part of
the COMBO-17 survey \citep{wolf01}.  The survey has imaged one
square degree of sky split over four fields using the Wide-Field
Imager (WFI) at the MPG/ESO 2.2m telescope on La Silla, Chile.  For
lensing studies, one of the fields was centered on the A901/A902
supercluster.  The filter set was chosen such as to provide reliable
classification and photometric redshift estimators, and consists of
five broad-band filters ($UBVRI$) and 12 narrow-band filters ranging
from 420 to 914 nm.  A deep $R$-band image taken under the best seeing
conditions during each run provides excellent data for gravitational
lensing studies.

In \S\ref{sec:pipeline}, we outline the initial processing applied to
images under the standard WFI pipeline reduction developed at the MPIA
Heidelberg. Although the resulting images are adequate for almost all
astrophysical applications, we need further higher-order corrections
for the detection of unbiased lensing shear. This is described in
\S\ref{sec:astrometry}.

The data used in this paper are the deep $R$-band and supplementary
$B$-band images from the COMBO-17 observations of the supercluster
field.  We will use the $R$-band image for a weak shear analysis of
the supercluster, in combination with color information from the
bluer band to separate foreground and background galaxy populations.
The details of the observations are presented in Table~\ref{tab:obs}.
The discussion of the full 17 filter dataset for the A901/902 field and
the photometric redshifts derived from it is reserved for a further
paper.

\begin{table*}
{\small
\begin{center}
\centerline{\sc Table 1}
\vspace{0.1cm}
\centerline{\sc Summary of Observations Used for the Present Study}
\vspace{0.3cm}
\begin{tabular}{lcrp{0.5cm}lcr}
\hline\hline
\noalign{\smallskip}
\multicolumn{1}{c}{Date} & {Filter} & {Exposures} & {} & \multicolumn{1}{c}{Date} & {Filter} &
{Exposures}\\
\hline
\noalign{\smallskip}
1999 Feb 18  &   B   &     10$\times$500 s & & 2000 Jan 31 &   R     & 8$\times$600 s\\
1999 Feb 19  &   B   &     12$\times$500 s & & 2000 Feb 06 &   R    & 27$\times$500 s\\
             &       &                     & & 2000 Feb 12  &   R    & 9$\times$500 s\\  
\cline{1-3}
\cline{5-7}
\noalign{\smallskip}
	&  \multicolumn{1}{r}{\bf total:} &   {\bf 3.05 h} & &
	&  \multicolumn{1}{r}{\bf total:} &  {\bf 6.33 h}\\
\noalign{\hrule}
\noalign{\smallskip}
\end{tabular}
\end{center}
}
\label{tab:obs}
\end{table*}

\subsection{Initial processing by standard WFI pipeline}\label{sec:pipeline}
Imaging by the Wide Field Imager (WFI) on the 2.2m telescope at La
Silla, Chile is carried out with a $4 \times 2$ array of $2048 \times
4096$ pixel CCDs.  For clarity and ease of reference, chips are
labelled by the letters `a', starting at the top left of the mosaic,
through a clockwise direction to chip `h' located at the bottom left
of the mosaic.  With a vertical chip-to-chip separation of 59 pixels
and a corresponding horizontal separation of 98 pixels, the scale of 1
pix $\equiv 0.238''$ gives a total field of view (FOV) for the WFI of
$0.56^{\circ} \times 0.55^{\circ}$.

De-biasing is conducted by individually subtracting the (vertically
smoothed) level measured in the overscan region of each chip image.
Non-linearity (i.e. departure from the proportionality between the
number of incident photons and the electric charge read out at the end
of an exposure, which worsens with increasing intensity), is corrected
by scaling each de-biassed chip image according to an ESO
laboratory-determined factor \citep{WFIman}\footnote{The WFI user manual is available via
the ESO homepage at {\tt http://www.eso.org}.}. Only chips b, c, f \&
g need be corrected in this way, the worst non-linearity being
exhibited by chip f which at saturation (65000 counts) deviates by
$1.4\%$ from linearity.

Production of mosaiced images must be handled with care due to the
physical complication of chip-to-chip misalignments in the CCD
array. Translational misalignments are easily accommodated, however
intrinsic chip rotations with respect to the array are not. A slight
chip rotation produces the added complexity of having to rotate the
chip image before insertion into the mosaic. 

Rather than rebin images, rotations are approximated by dividing up
the misaligned chip images into horizontal strips. Each strip is then
inserted into a temporary chip image with a horizontal offset
determined by the required rotation. This temporary chip image is then
divided into vertical strips and then finally inserted into the mosaic
with vertical offsets according to the rotation.  Adopting this
shearing method considerably reduces data reduction time, is optimal
for cosmic ray removal (see below), and has been shown to have no
effect on photometry. Table 2 lists chip rotations determined
independently in this work as part of the astrometric fitting
described in \S\ref{sec:linast}, although they are in agreement with
the rotations incorporated in the WFI pipeline.

After producing mosaic images in this fashion, science frames were
flattened with normalized mosaic flat-field frames produced in exactly
the same way.  Since each physical pixel is maintained intact
throughout the reduction process, cosmic ray hits and cosmetic chip
defects can be detected very efficiently by comparing dithered images
taken through the same filter. To this end, flat-field corrected
mosaic images are aligned with respect to moderately bright stars in
the field (to integer pixel accuracy). The cosmic ray detection
algorithm employs a $\sigma$-clipping with respect to the
median pixel value derived from at least five exposures with
comparable seeing. Discrepant values are replaced by the median value.

\subsection{Astrometry}\label{sec:astrometry}

\subsubsection{Linear astrometric fitting}\label{sec:linast}
While the creation of mosaiced images by the standard reduction
pipeline described above is sufficient for most purposes, we require a
more rigorous characterization of the geometry of the camera for our
weak shear study.  For example, intrinsic rotation of the FOV of the
WFI is known to be caused by imperfect alignment of the 2.2m telescope
axis of rotation with the celestial pole as well as telescope flexure
in tracking across the sky. Other causes are atmospheric in origin,
due to the differential refraction of object images by the Earth's
atmosphere and the fact that an offset in right ascension produces a
rotation due to the non-perpendicularity of lines of declination away
from the celestial equator.

To this end, we chose to treat each CCD image as an individual
contribution to the final coadded image, and to compute an astrometric
solution separately for each chip image.  For each debiased,
normalized, flattened and cosmic-ray cleaned mosaic the approximate
rotations described above were removed to restore the original
detector coordinate system.  Next, the eight component chip images
were extracted from each mosaic.  Henceforth the term `image' shall
refer to an individual 2K$\times$4K chip image.

Using SExtractor 2.1  \citep{bertin96} an initial catalogue of
bright objects was created for each image.  A rough transformation was
performed to convert the pixel coordinates ($x,y$) to celestial
coordinates ($\alpha,\delta$) using the pointing information encoded
in the image header.  The objects in the catalogue were then matched
to the SuperCOSMOS Southern Sky Survey \footnote{\texttt{http://www-wfau.roe.ac.uk/sss}}
within a tolerance of 5\arcsec.  These reference objects were used to
iteratively calculate a linear astrometric solution for the image,
with the tangent point for projection being the optical axis.  The rms
residuals of the objects used in the final solution (typically
numbering $\sim$400 per image) differed from the median by less than
3$\sigma$.

The linear fit used takes the form:
\begin{equation}
\begin{array}{rcl}
            x  &=&  a_0 + a_1\xi + a_2\eta \\
            y  &=&  a_3 + a_4\xi + a_5\eta,
\end{array}
\end{equation}
where the `expected' coordinates (standard coordinates on the tangent
plane, projected about the optical axis) are related to the `measured'
coordinates (x-y coordinates on the detector plane) by a 6 coefficient
linear fit.  Note that field-to-field variations were found to be
significant, so the external calibration was calculated {\em for each
image} individually.

The linear fit can be decomposed into terms such as pixel scale, rotation,
and non-perpendicularity of the coordinate axes for each image.  The
average rotation angles required to align the x-y axes to a N-E
orientation and the angle of deviation from the perpendicular for the
x-y axes are listed in Table~\ref{tab:rot} for each chip.  The
resulting fits produced median rms residuals of 0.2 arcsec, i.e. less than 1
pixel and close to the limiting accuracy of the photographic data.

\begin{table*}
{\small
\begin{center}
\centerline{\sc Table 2}
\vspace{0.1cm}
\centerline{\sc Decomposition of the Astrometric Fits}
\vspace{0.3cm}
\begin{tabular}{cr@{.}lrrrrr@{.}l}
\hline\hline
\noalign{\smallskip}
 & \multicolumn{5}{c}{Linear Fit}  & & \multicolumn{2}{c}{Radial Fit}\\
\cline{2-6}
\cline{8-9}
\noalign{\smallskip}
Chip & \multicolumn{2}{c}{Rotation} & \multicolumn{1}{c}{$\pm$} & Skew
& \multicolumn{1}{c}{$\pm$} & &\multicolumn{2}{c}{$C$}\\
\hline
\noalign{\smallskip}
 a   & -288&3 &   53.5 & -104.5 &   21.4 & \ \ \ \ \ \ \ & 6&4\\
 b   &  276&0 &   54.0 & -117.0 &   17.2 & \ \ \ \ \ \ \ & -38&7\\
 c   &  126&0 &   53.2 &  120.1 &   20.2 & \ \ \ \ \ \ \ & -29&4\\
 d   &  130&7 &   54.5 &  161.9 &   19.8 & \ \ \ \ \ \ \ & 0&7\\
 e   &  144&9 &   53.9 & -157.3 &   11.6 & \ \ \ \ \ \ \ & 1&6\\
 f   &  123&8 &   53.0 & -115.2 &   16.3 & \ \ \ \ \ \ \ & -39&8\\
 g   &  120&9 &   53.5 &  104.8 &   24.4 & \ \ \ \ \ \ \ & -50&6\\
 h   &  -98&3 &   49.9 &   74.7 &   27.2 & \ \ \ \ \ \ \ &  13&0\\
\noalign{\hrule}
\noalign{\smallskip}
\vspace{-1.5cm}
\tablecomments{Columns 2-5 list the angles of rotation and
non-perpendicularity (in arcseconds) for the eight chips according to the
linear astrometric fit.  The final column lists the radial distortion
coefficient (radians$^{-2}$) found when a radial distortion term (as in
eq.~[\ref{eqn:rad}]) was added to the fit.}
\end{tabular}
\end{center}
}
\label{tab:rot}
\end{table*}

\subsubsection{Quantifying shear induced by radial distortions}
While the linear fits produced satisfactory results, it is
nevertheless important to consider higher-order or radial distortions.
As a test to search for pincushion or barrel distortions, a radial
distortion term was added to the linear fit, of the form:
\begin{equation}\label{eqn:rad}
    r^\prime = r(1 + Cr^2),
\end{equation}
where $r$ is the radial distance from the tangent point (optical
axis), $c$ is the distortion coefficient, and $r^\prime$ is the radial
distance from the tangent point in the presence of the distortion.

The fit for some chips (particularly the y-component of the fit for
the inner four chips) improved somewhat when the radial term was
added, but there was no significant improvement in the resulting rms
residuals for the instrument as a whole.  However, the results of the
radial fitting do allow us to constrain the amount of linear
distortion found in each chip.  The resulting values of $C$ (listed in
the final column Table~\ref{tab:rot}) were small, and had a small
dispersion for each chip from image to image.  The contrast between the
values for the inner and outer four chips indicates that the
mosaic as a whole does not display a typical radial distortion
pattern.

Knowing the distance of the edge of each chip from the tangent point,
this yields distortions of $\delta r/r \sim 0.025$\% at the furthest
corners of the camera, or slightly higher for the inner chips.  This
tallies well with the figures quoted in the WFI manual, which claims
geometric distortions $\leq$0.08\% across the entire camera.  

We can further quantify any artificial shear induced by such a radial
distortion using the relations of \cite{bacon00}.  If the
displacement $\delta {\bf r} = Cr^3$ and ${\bf \hat{r}}$ is the unit
radial vector, then the induced instrumental shear is $\gamma_i = Cr^2
\hat{e}_i$, where $\hat{e}_i$ is the unit radial ellipticity vector.
Using the radial distortion coefficients derived from our astrometric
fits, we find a shear pattern with amplitude $\gamma < 0.0001$
throughout.

It is quite clear that despite the wide field of view of the camera,
the instrumental distortion in the WFI is extremely small and will not
significantly affect our shear measurements.  We shall continue with
the analysis using the simple linear fit with no correction for the
negligible radial distortion. Note that corrections involving
higher-order polynomials could be used to remove any non-linear,
non-radial distortions present, but we see little evidence of these
from the residuals following the linear fit.

\begin{figure*}
\epsfig{file=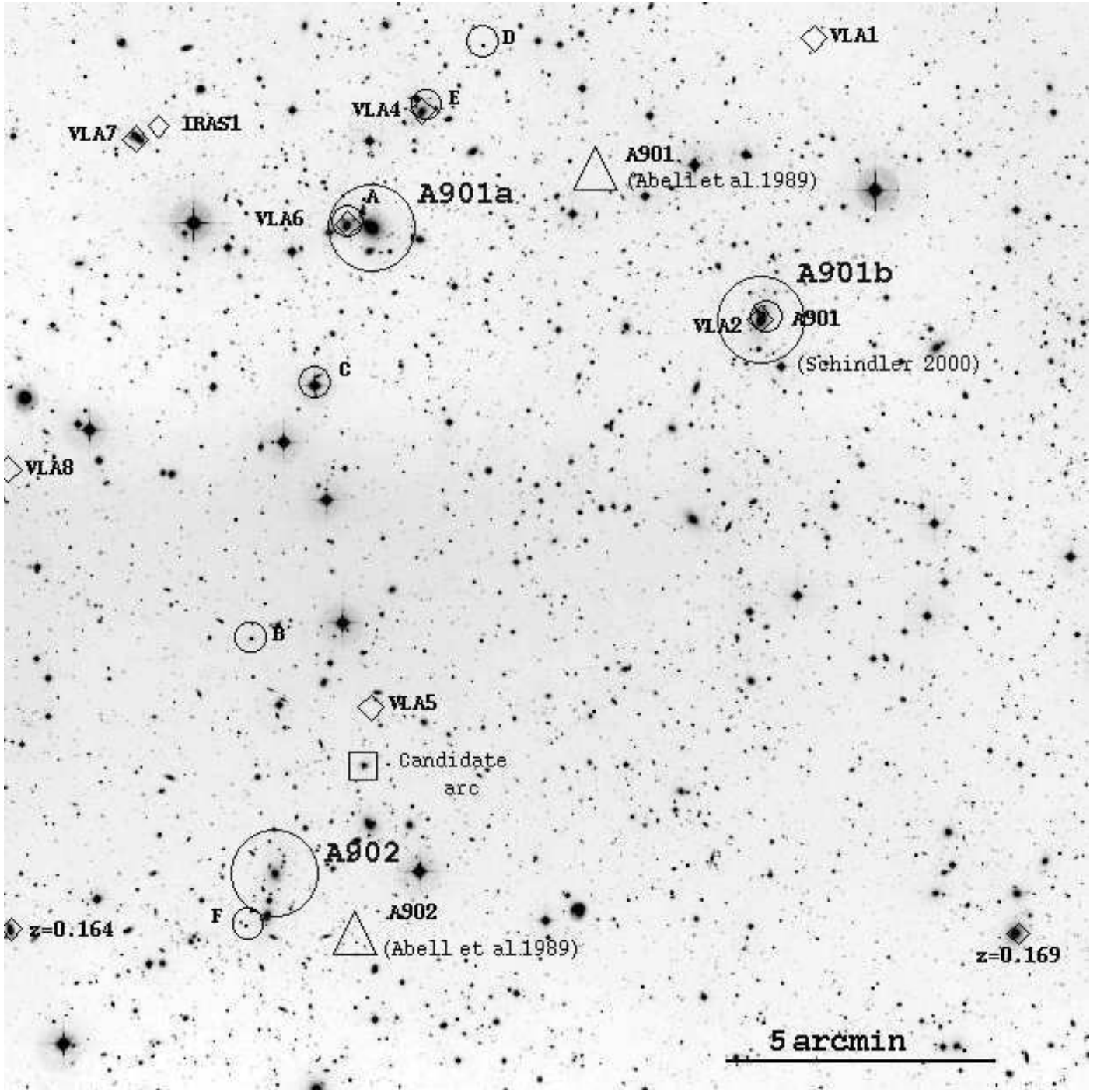,width=\textwidth}
\caption{The $R$-band view of the central $20\times20$ arcmin of the
supercluster field (seen at reduced resolution).  The two large
triangles mark the positions of Abell 901 and Abell 902 in the
original catalogue of galaxy clusters \citep{aco89}, while the
three large circles indicate the optical centers of the clusters that
we have adopted in this paper.  The smaller circles indicate the
locations of the ROSAT/HRI X-ray sources from \cite{schindler00}.  The
remaining objects marked by diamonds are two galaxies with known
redshift  \citep[also from][]{schindler00}, radio detections from the
NRAO/VLA All Sky Survey  \citep[][ labelled `VLA']{condon98} and the
IRAS Faint Source Catalogue  \citep[][labelled `IRAS']{moshir90}.
North is up, east is to the left.}
\label{fig:rbandim}
\end{figure*}

\begin{figure}
\centerline{\epsfig{file=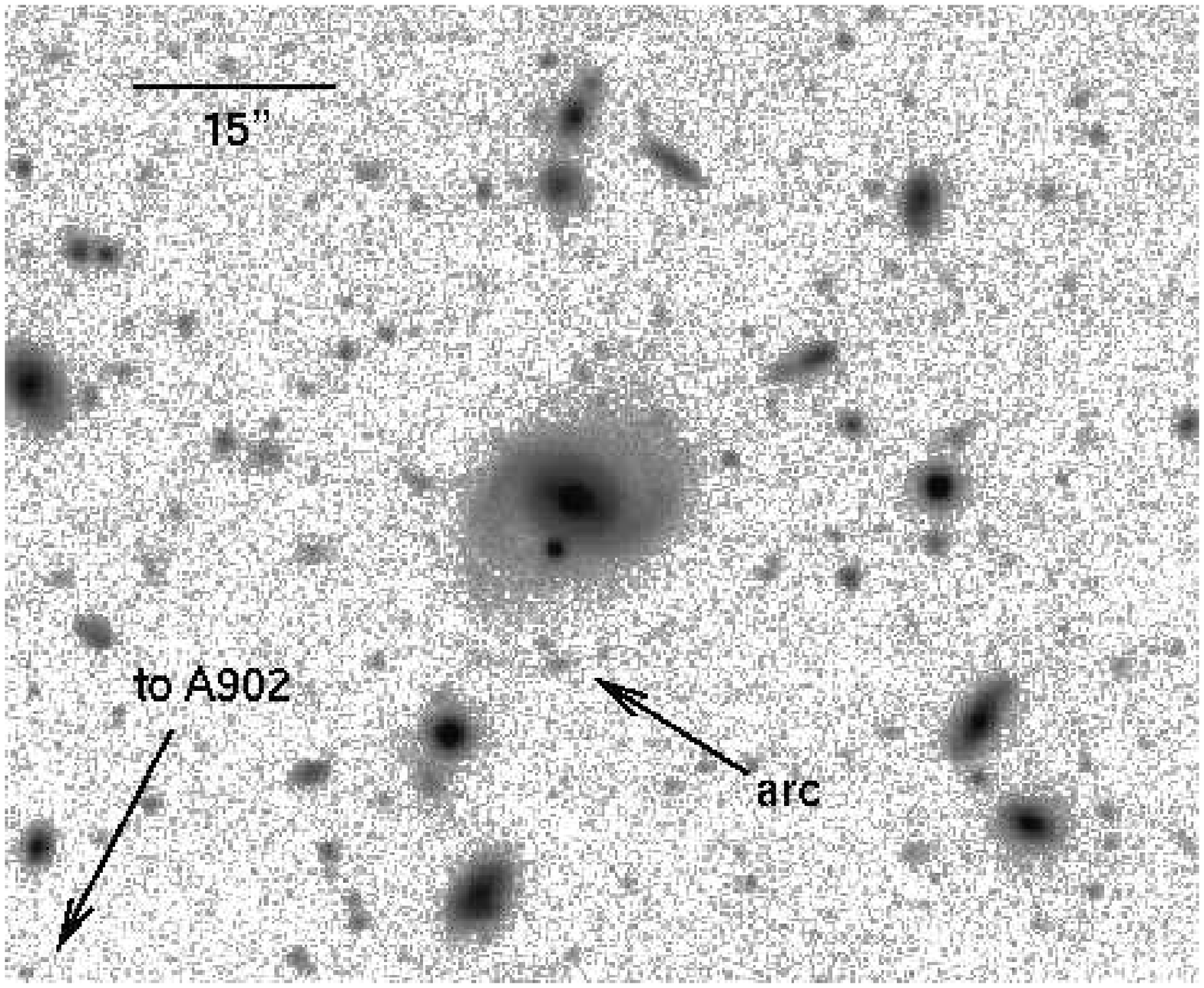,width=0.45\textwidth}}
\caption{Candidate giant arc.  The potential deflecting galaxy is a barred
spiral with $R=18$ and is located $\sim$2.5 arcmin from the center of
Abell 902.}
\label{fig:arc}
\end{figure}

\subsubsection{Registration and coaddition}
The linear astrometric fits for each individual chip image were used
to register the images to the same coordinate system.  The images were
thus aligned and combined with 3$\sigma$ clipping rejection of bad
pixels, scaling by the median and weighting by exposure time.  Due to
the large number of images contributing to each pixel in the final
image (44 for the $R$-band, 22 for the $B$-band), most bad columns and
pixels were removed by the $\sigma$-clipping algorithm.  The combined
mosaiced images were trimmed of the noisy outlying regions where the
overlap from the dithering pattern was not complete.  The resulting
final images used for the weak shear analysis measured $8192 \times
8192$ pixels ($32.5\arcmin \times 32.5\arcmin$), and the measured seeing
was 0.7\arcsec\  ($R$-band) and 1.1\arcsec\  ($B$-band).

Fig.~\ref{fig:rbandim} shows the inner regions of the
$R$-band image, along with the published positions of the clusters
from \cite{aco89} and the X-ray sources from \cite{schindler00}.
In addition, Fig.~\ref{fig:arc} shows a candidate strongly lensed
source.  The arc-like structure is located 13 arcsec from a bright
barred spiral galaxy, which is in turn located $\sim$2.5 arcmin from
the adopted center of Abell 902.  The deflecting galaxy has $R=18$ and
$B-R$ color similar to the cluster galaxies.

\subsection{Photometric calibration}

Photometric calibration was provided from two spectrophotometric
COMBO-17 standards \citep[see]{wolf01} in the A901/902 field
selected from the spectral data\-base of the Hamburg/ESO (HE) Survey
\citep{wisotzki00}.  Spectra were obtained using a wide ($5\arcsec$)
slit on the ESO/Danish 1.54m and the ESO 1.52m telescope at La Silla
in November 2000. Correcting the observed A901/902 standard spectra to
remove the instrument, optics and atmosphere signature, photometric
scalings were determined from the HE survey standards. Using the
$B$- and $R$-band WFI filter responses, zero points were calculated
and magnitudes transformed to the Vega system.  Apparent magnitudes
were corrected for extinction using the IRAS dust maps of
\cite{schlegel98}.  For an average $E(B-V)=0.058$ and assuming an $R_V
= 3.1$ extinction curve, we derive corrections $\Delta m_B = 0.25$ and
$\Delta m_R = 0.16$.  We estimate that the $S/N=3$ limiting magnitude
in the two bands is $R=25.7$ and $B=25.5$.

\section{Point spread function corrections}\label{sec:psf}

In this section we outline the correction for such effects as variable
seeing conditions from exposure to exposure, telescope tracking
errors, as well as smearing of objects due to imperfect alignment of
images before coaddition and the circularization of small objects by
the seeing.  These corrections have been discussed in detail in
several previous papers \citep*[hereafter KSB, with refinements in
\citealt{lk97} and \citealt{hoekstra98}]{ksb}.

\subsection{Object catalogues}
The {\tt imcat} software described in \cite{kaiser99} was used to
determine the shape parameters for the faint galaxy sample used in the
weak lensing analysis.  For ease of computing, the {\tt hfindpeaks}
object detection routine was performed on 2K$\times$2K image sections,
and the resulting catalogues shifted by the appropriate amount and
concatenated together.  The local sky background was estimated using
{\tt getsky} and and aperture magnitudes and half-light radii, $r_h$,
for each object were calculated using the {\tt apphot} routine.
Finally, {\tt getshapes} was used to calculate the weighted quadrupole
moments, defined as 
\begin{equation}
I_{ij} \equiv \int d^2x\ w(x) x_i x_j I(x),
\end{equation}
where $I(x)$ is the surface brightness of the object at angular
position $x$ from the object center, and $w(x)$ is a Gaussian weight
function.  The scale length, $r_g$, of the weight function is
previously determined by {\tt hfindpeaks} as the radius of the Mexican
hat filter function that maximizes the signal-to-noise of the object
detection.  Finally, the weighted quadrupole moments are used to
calculate the ellipticity components
\begin{equation}
e_1 = \frac{I_{11}-I_{22}}{I_{11}+I_{22}}, \hspace{1cm} e_2 = \frac{2I_{21}}{I_{11}+I_{22}}.
\end{equation}

Photometric catalogues were also constructed for each image, using
SExtractor 2.1.  The detection criteria were defined such that an
object was required to be $1.5\sigma$ above the background and
comprising at least 7 connected pixels.  The photometric information
(from SExtractor) and the shape estimates (from {\tt imcat}) were then
merged to provide the final catalogue for lensing analysis,
conservatively requiring that an object be detected by both software
packages.  Due to the superior seeing and longer exposure time, the
shape parameters derived from the $R$-band image will be used for the
weak shear reconstructions, although as we describe in
$\S\ref{sec-col}$, the $B-R$ colors will be used to discriminate
between the foreground (including the cluster galaxies) and the
background populations.

Finally, a mask was created directly from the image to remove objects from the
catalogue that lay in areas contaminated by bright stars, ghost
images, and diffraction spikes.  This masked region totalled only
3.4\% of the total area of the image, and can be seen in
Fig.~\ref{fig:fieldplot}.

\subsection{Anisotropic PSF correction}

\begin{figure}
\epsfig{file=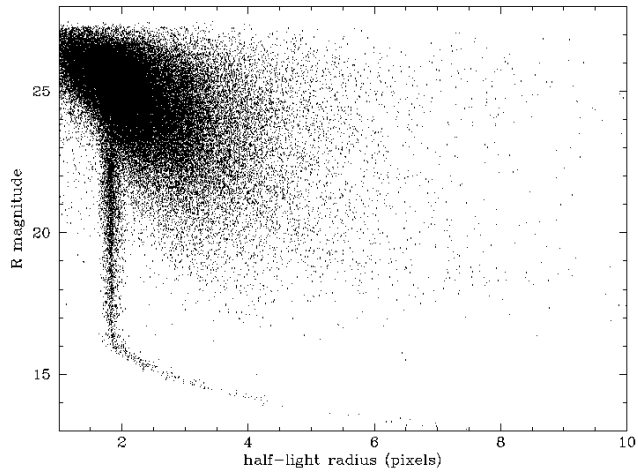,height=0.5\textwidth,angle=270}
\caption{Magnitude vs. half-light radius of all objects in the initial
$R$-band catalogue.  Stars form a clear vertical locus and can be
easily differentiated from galaxies down to $R\sim 23$. Saturated
stars deviate from this column towards larger half-light radii at
bright magnitudes.}
\label{fig:rhmag}
\end{figure}

Fig.~\ref{fig:rhmag} shows the size-magnitude diagram for the full
sample of objects.  The stars are clearly visible as a column of
objects with half-light radii of 1.8 pixels, and can be distinguished
from galaxies down to $R\sim 23$.  To examine the behavior of the
point-spread function (PSF) across the image we select a sample of
non-saturated ($R>16$) stars.

\begin{figure*}
\centerline{\epsfig{file=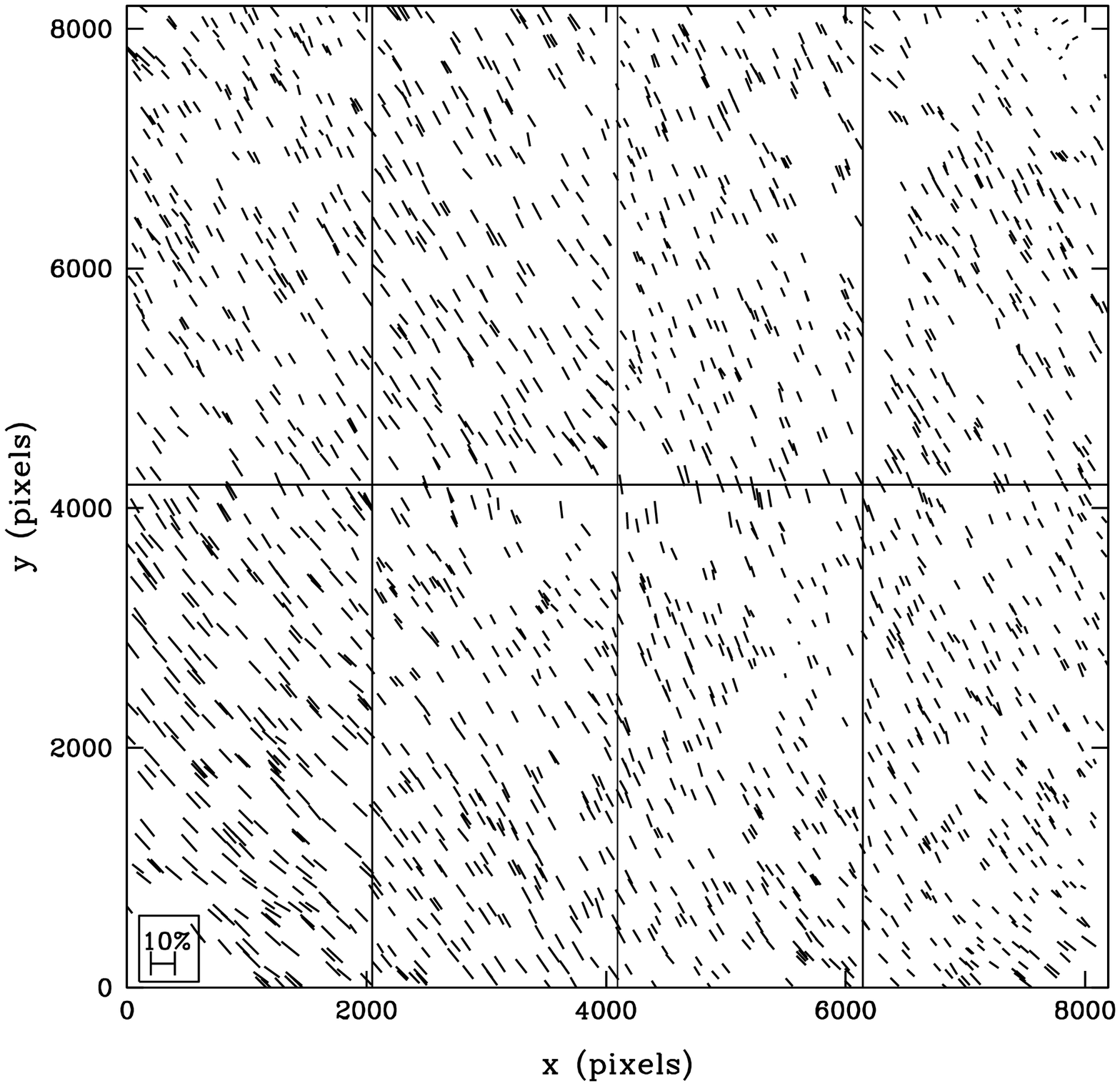,width=0.5\textwidth}
            \epsfig{file=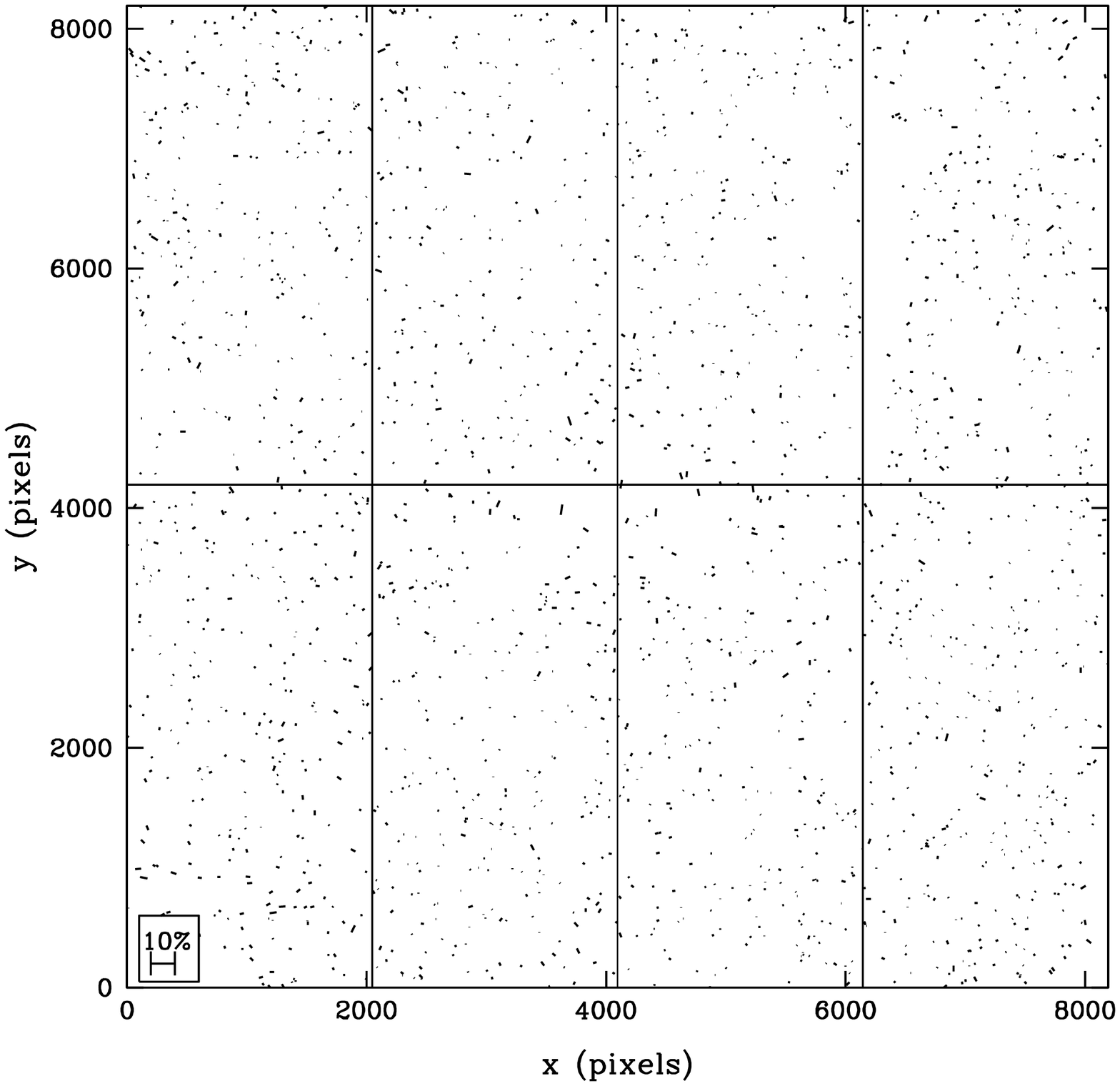,width=0.5\textwidth}}
\caption{Ellipticity pattern of stars on the $R$-band coadded image,
before ({\em left}) and after ({\em right}) correction for the
anisotropic PSF.  The horizontal and vertical lines show the
approximate chip divisions used for the polynomial fitting. }
\label{fig:starsticks}
\end{figure*}

\begin{figure}
\centerline{\epsfig{file=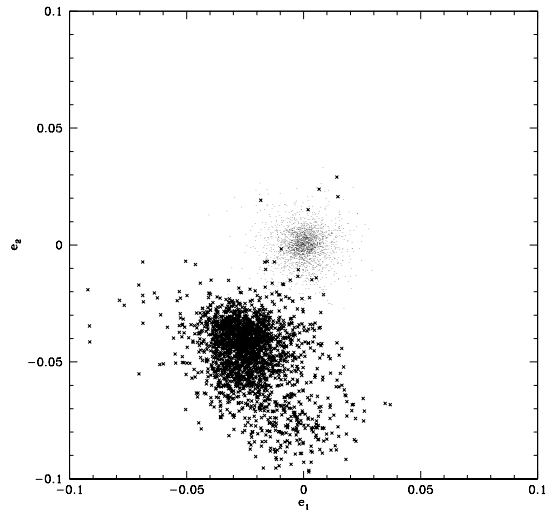,width=0.4\textwidth}}
\caption{Ellipticities distribution of stars, before (crosses) and after (dots)
correction for PSF anisotropies.}
\label{fig:estars}
\end{figure}

The stellar ellipticity pattern across the final summed $R$-band image
is shown in Fig.~\ref{fig:starsticks}, and the distribution of
ellipticities in Fig.~\ref{fig:estars}.  While there is overlap
between the chip regions due to the dithering pattern, we divide the
mosaiced image into 8 approximate chip regions for the purposes of the
PSF correction.  Within these regions, outlined in
Fig.~\ref{fig:starsticks}, the PSF is smoothly varying and even within
the overlap regions there are no sharp discontinuities in the
behavior of the PSF.

We apply preliminary cuts to our catalogue to exclude objects
unsuitable for shear analysis.  We remove those objects with a
half-light radius smaller than the stellar half-light radius, as well
as objects with an {\tt imcat} significance $\nu<5$ and with
ellipticity $e>0.5$.  Next, within each chip region we fitted a
two-dimensional cubic to the measured stellar ellipticities, iterating twice to
remove outliers with large residuals.  

The modelled stellar ellipticities were used to correct the
ellipticities of the galaxies by

\begin{equation}
e_{\rm corr} = e - \frac{P_{\rm sm}}{P_{\rm sm}^{*}} e^{*},
\end{equation}
where starred quantities referred to stellar properties, and $P_{\rm
sm}$ is the {\tt imcat} `smear polarizability' matrix of higher
moments of surface brightness described in KSB.  As the non-diagonal
elements are small ($< 2 \times 10^{-4}$) and the diagonal elements
nearly equal, we choose to treat this as a scalar equal to half the
trace.  The mean stellar ellipticity before correction and after
correction was then
\[ \begin{array}{llc}
 e_1 = 0.014 \pm 0.011 & e_2 = -0.047 \pm 0.014 & {\rm (before)}\\
 e_1 = 0.00013 \pm 0.006 &e_2 = 0.00003 \pm 0.006 & {\rm (after)}.
\end{array} \]
Fig.~\ref{fig:starsticks} and Fig.~\ref{fig:estars} show the
resulting pattern and distribution of corrected stellar ellipticities.

\subsection{Isotropic PSF correction}

The isotropic correction recovers the lensing shear prior to
circularization by the PSF.  Following \cite{lk97} and
\cite{hoekstra98} we use the corrected galaxy ellipticities in
conjunction with the `shear polarizability' tensor $P_{\rm sh}$ to
calculate the shear estimate
\begin{equation}\label{eqn:pgamma}
\gamma = P^{-1}_{\gamma} e_{\rm corr},
\end{equation}
where
\begin{equation}
P_{\gamma} = P_{\rm sh} - \frac{P_{\rm sh}^{*}}{P_{\rm sm}^{*}} P_{\rm sm}.
\end{equation}
Again, the starred quantities refer to stellar properties, and $P_{\rm
sh}^{*}$ and $P_{\rm sm}^{*}$ are treated as scalars equal to half the
trace of the respective matrices.  To calculate $P_\gamma$ for the
galaxies, we fit $P^{11}_{\gamma}$ and $P^{22}_{\gamma}$ as a function
of smoothing radius $r_g$, and insert the fitted value into
equation~(\ref{eqn:pgamma}) to calculate the shear measurement for each
galaxy.  Finally, we remove from the catalogue any galaxies with a
resulting $|\gamma| > 2$ due to noise.  Note that this procedure
actually yields the {\em reduced shear}, $\gamma/(1-\kappa)$, but in
the wide-field, weak lensing limit ($\kappa \ll 1, \gamma \ll 1$) in
which we are working this reduces to the shear, $\gamma$.

\section{Cluster properties}\label{sec:clusters}

To identify the likely members of the supercluster, we first spatially
separate
the catalogue of galaxies.  We then use the resulting samples to isolate the
tight sequence of early-type supercluster galaxies on the $B-R$ vs. $R$
color-magnitude diagram.

\begin{figure*}
\centerline{\epsfig{file=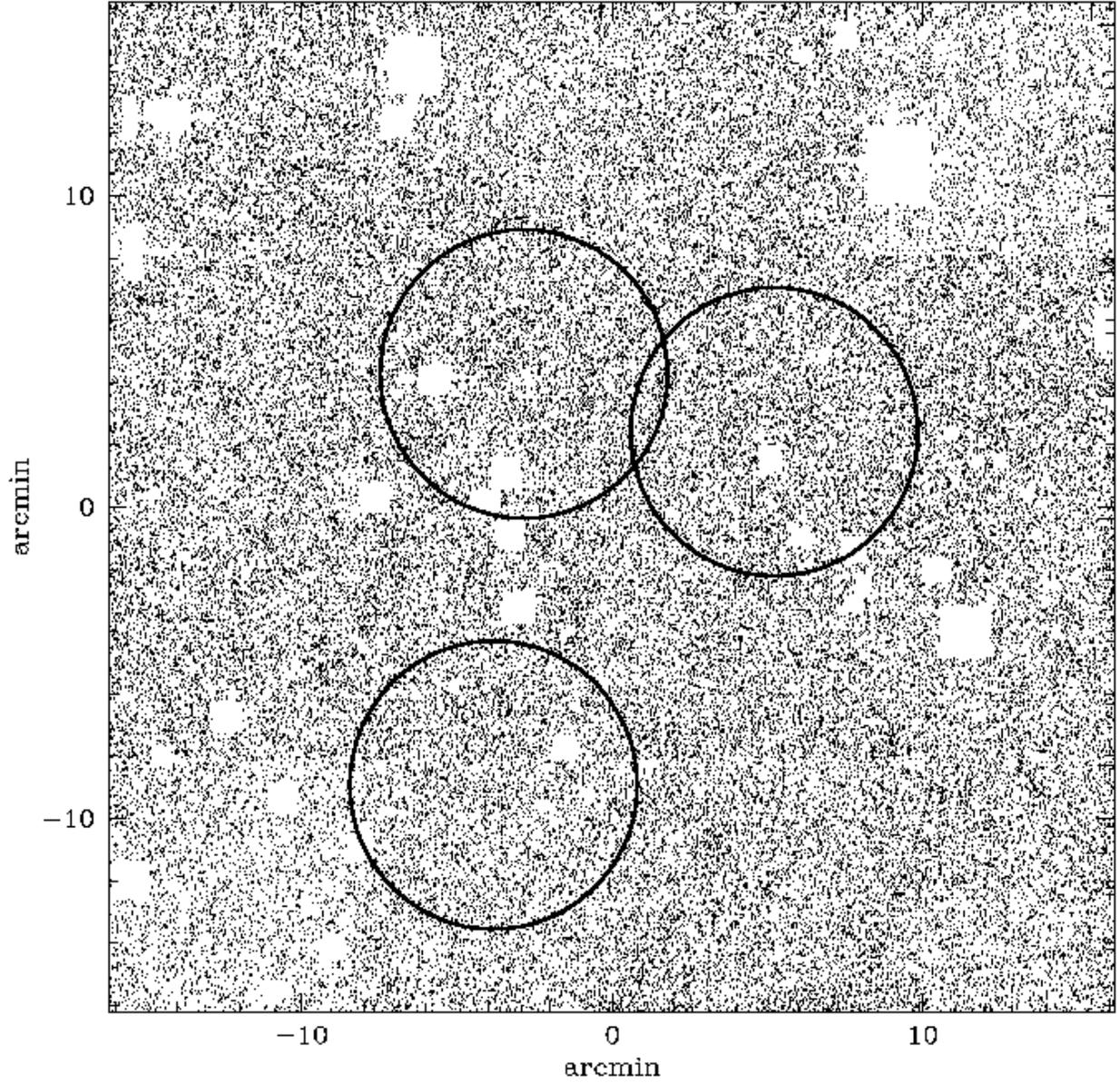,width=0.9\textwidth}}
\caption{Division of catalogue into `cluster' and `field' regions.
Each dot represents one galaxy in the catalogue, with the area masked
out due to contamination by bright stars and diffraction spikes
indicated by the blank rectangular regions.  The circles are apertures
of radius 4.6 arcmin (500$h^{-1}$ kpc) around the optical center of
each cluster and define the `cluster' regions used to isolate the
color-magnitude sequence of supercluster galaxies
(cf. Fig.~\ref{fig:numbercounts} and
Fig.~\ref{fig:colmagfieldclus}).}
\label{fig:fieldplot}
\end{figure*}

Fig.~\ref{fig:fieldplot} shows the locations of all the objects in the
catalogue within the $32.5\arcmin\times 32.5\arcmin$ field of view.
The large circles represent an aperture of radius 4.6 arcmin around each
cluster.  The angular size scales as
\begin{equation}
R(\theta)  = 0.87 D_A(z_L) (\theta/1^{\prime}) h^{-1} {\rm Mpc}
\end{equation}
where the dimensionless angular distance
\begin{eqnarray}
D_A(z) &=& \frac{1}{(1+z)} \int^z \frac{dz}{[\Omega_m(1+z)^3 +
            \Omega_\Lambda]^{1/2}},\\
                &\simeq& \frac{z}{(1+z)(1+3/4 \Omega_m z)}
\end{eqnarray}
and the second line is a good approximation to the low-redshift
angular distance in a spatially flat model with cosmological
constant.  For our supercluster at $z=0.16$ this is relatively
insensitive to cosmology, and yields 
\begin{eqnarray}
R(\theta) &=& 0.11 (\theta/1^{\prime})h^{-1} {\rm Mpc} \\
          &=& 0.002 (\theta/1^{\prime\prime})h^{-1} {\rm Mpc}.
\end{eqnarray}
Thus the radii of the apertures in Fig.~\ref{fig:fieldplot} correspond to
0.5$h^{-1}$ Mpc at the redshift of the supercluster.

Table~\ref{tab:pos} lists the coordinates chosen as the center of each
of the three clusters: in Abell 901a and 901b this corresponded to the
obvious brightest cluster galaxy of \citep[close to the X-ray
positions of][]{schindler00}, and in Abell 902 the more northern of the
two bright galaxies in the center of overdensity of numbers in that
region.  Note that these three positions differ slightly from the
(two) positions given in \cite{aco89} for Abell 901 and Abell 902 (see
Fig~\ref{fig:rbandim}).

An attempt to quantify the richness of each of the three clusters is
hampered by their close proximity.  The original Abell richness class
\citep{abell1958} is determined from the number of galaxies in the
magnitude range $m_3$ to $m_3+2$ (where $m_3$ is the magnitude of the
third brightest cluster galaxy) contained within a 1.5$h^{-1}$ Mpc
radius.  In the case of the A901/902 supercluster, the maximum
separation between Abell 901a and Abell 902 is only 7.8 arcmin, or
$\sim$850 $h^{-1}$ kpc at $z=0.16$.  We can therefore at best place a
lower limit on the Abell richness class by considering the counts
within a 3.9 arcmin radius.  This yields $N_{\rm Abell} = 34, 27,$ and
$20$ galaxies for A901a, A901b, and A902, respectively, which
corresponds to an Abell richness class 0 in each case.  
More informative, however, is to consider the supercluster as a single
system.  In this case we can probe the entire $R_{\rm Abell} = 1.5
h^{-1}$ Mpc (measuring from the centroid of the cluster positions) and
find $N_{\rm Abell}$ = 92, corresponding to an Abell richness class II.

\subsection{Color selection of early type galaxies}

\begin{figure}
\centerline{\epsfig{file=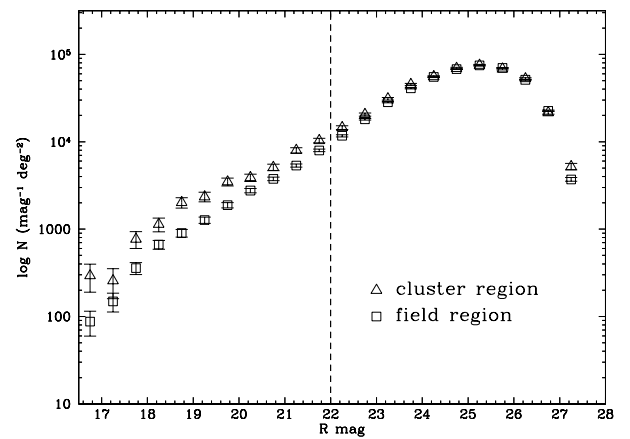,height=0.45\textwidth,angle=270}}
\caption{Number-magnitude relation for `cluster' and `field' regions
as defined in Fig.~\ref{fig:fieldplot}.  A clear excess of bright
galaxies in the cluster apertures is visible up to $R\sim 22$.}
\label{fig:numbercounts}
\end{figure}

\begin{figure*}
\centerline{\epsfig{file=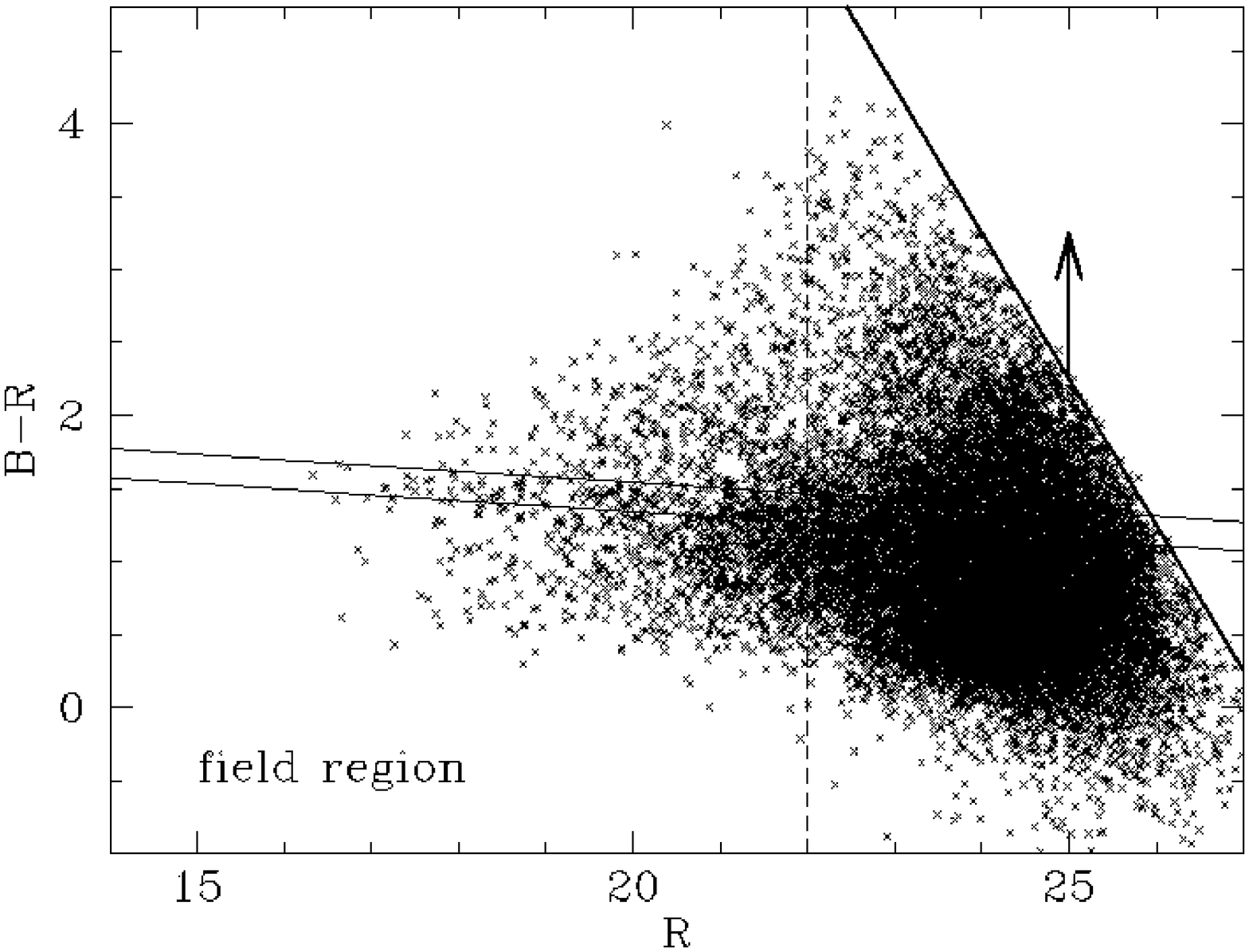,height=0.45\textwidth,angle=270}
            \epsfig{file=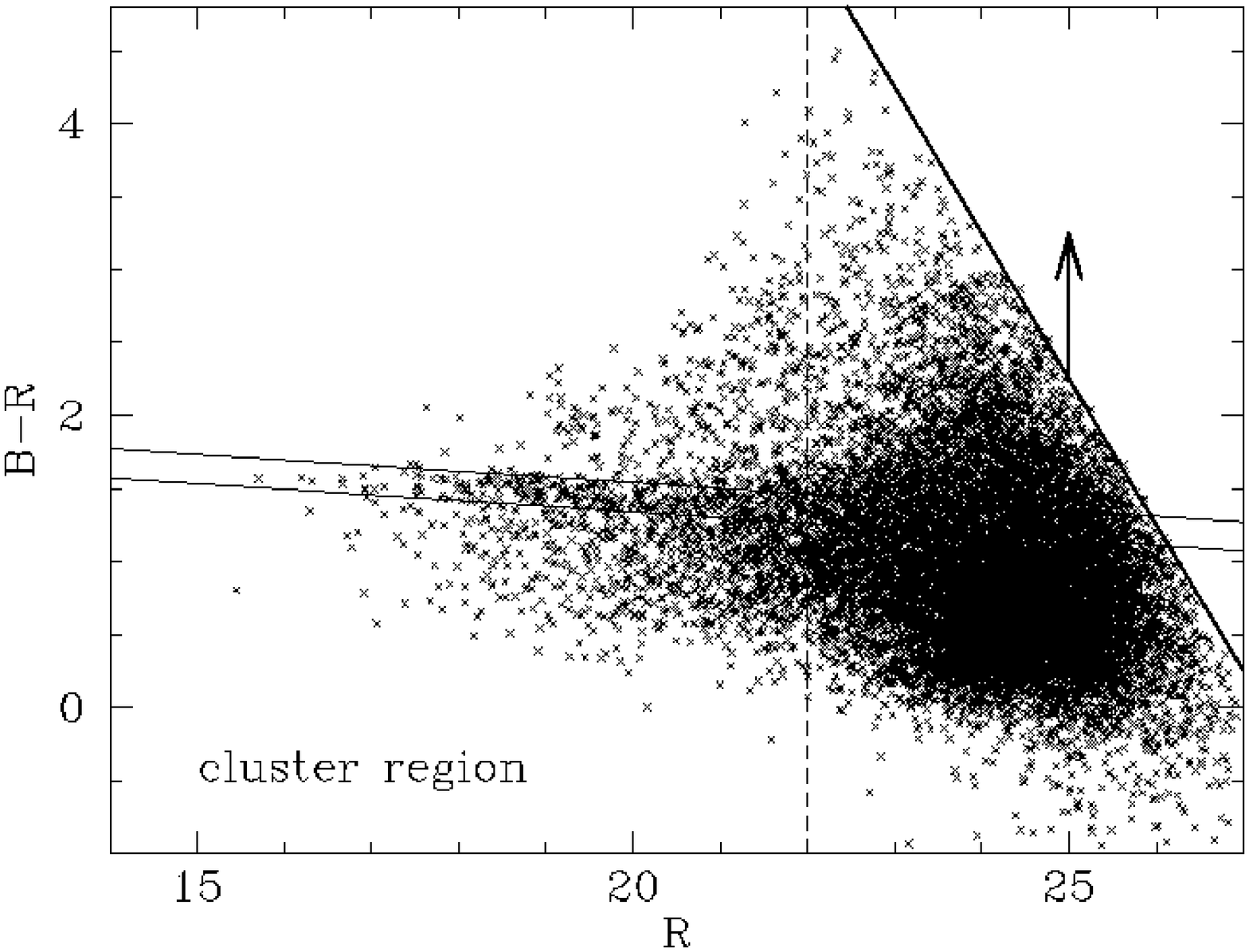,height=0.45\textwidth,angle=270}}
\caption{The $B-R$ vs. $R$ color-magnitude diagrams for the `field'
region (left) and the `cluster' region (right) as defined by the
boundaries in Fig.~\ref{fig:fieldplot}.  The early-type cluster
galaxies are clearly visible as a tight red sequence in the right-hand
plot.  The color selection criteria (eq.~[\ref{eqn:colcut}]) used to
isolate the supercluster galaxies is shown by the parallel lines.  Due
to the relative shallowness of the $B$-band image with respect to the
deep $R$-band image, approximately half the galaxies in each sample
are detected in the $R$-band alone.  The lower limit on the resulting
$B-R$ color for these galaxies is indicated by the thick line and
arrow.  The dashed line at $R=22$ serves to further subdivide the
populations into `bright' and `faint' samples.
}\label{fig:colmagfieldclus}
\end{figure*}

Dividing the catalogue spatially into `cluster' and `field' regions
according to boundaries shown in Fig.~\ref{fig:fieldplot}, we plot the
$R$-band number counts in Fig.~\ref{fig:numbercounts}.  A clear excess
xof bright galaxies in the cluster region is visible.  The $B-R$
vs. $R$ color-magnitude diagrams for the two regions
(Fig.~\ref{fig:colmagfieldclus}) reveal a prominent sequence of red
galaxies with small scatter in the region of the clusters.  We fit
this relation according to
\begin{equation}\label{eqn:colcut}
\left| B-R - (-0.0385R + 2.21)\right| < 0.1
\end{equation}
and will use this color-cut to separate the supercluster members from the
foreground and background population throughout the whole field of view,
using the entire catalogue of galaxies.

\begin{figure*}
\centerline{\epsfig{file=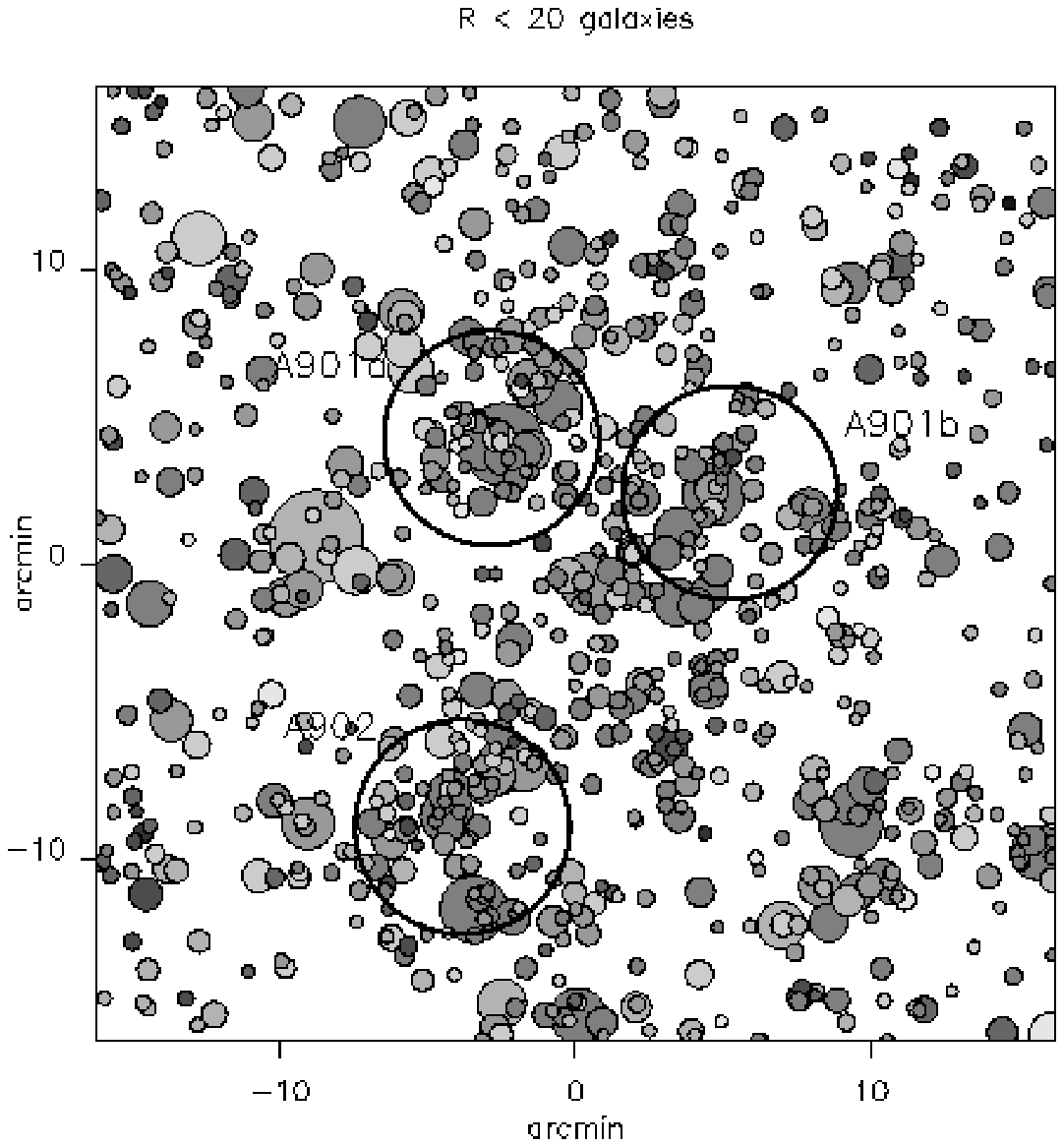,width=0.5\textwidth}
            \epsfig{file=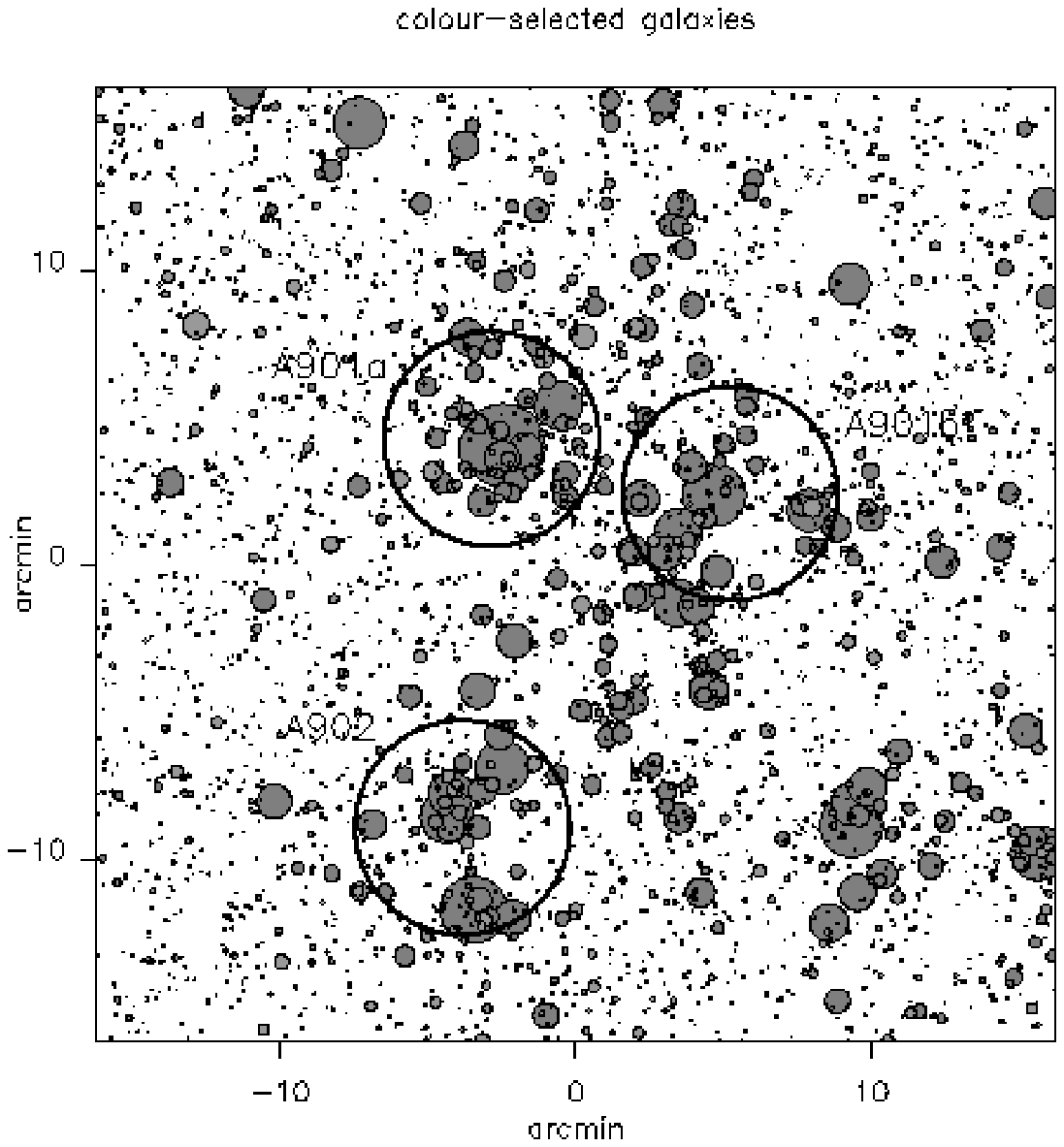,width=0.5\textwidth}}
\caption{Spatial distribution of galaxies. Left: Position of all
bright galaxies with $R<20$ in the supercluster field.  The area of
the symbol is proportional to the flux of the galaxy and a darker
shade indicates a redder $B-R$ color.  The large circles mark the
optical centers of the three Abell clusters as in
Fig.~\ref{fig:fieldplot}.   Right: The same, but restricted to
candidate supercluster members selected from the color-magnitude
diagram (Fig.~\ref{fig:colmagfieldclus}). This reveals not only that
the three main clusters have similar colors (indicating similar
redshifts) but also that the structure traced by the color-selected
galaxies extends further afield.  North is up, east is to the left.}
\label{fig:galaxies}
\end{figure*}

Fig.~\ref{fig:galaxies} shows the result of this color cut when it is
applied to the entire catalogue.  The left panel shows the
distribution of all the bright $(R <20)$ galaxies in the field, and
the prominent clustering present is clearly visible.  When the
color-cut of equation~(\ref{eqn:colcut}) is applied, the structure of the
supercluster as traced by the early-type galaxies is revealed.  The
distribution of these color-selected galaxies is shown in the right
panel of Fig.~\ref{fig:galaxies}.  The three clusters, sharing the
same color-magnitude relation for their early-type galaxies, are
clearly present.  Furthermore, the larger supercluster structure
becomes clear (including galaxies stretching between the clusters, and
an obvious sub-clump of supercluster galaxies approximately 12 arcmin
west of Abell 902).  This excess of supercluster galaxies outside the
cluster boundaries defined by the apertures above accounts for a
similar (but much weaker) color-sequence also visible in the
color-magnitude diagram for the `field' sample (left-hand side of
Fig.~\ref{fig:colmagfieldclus}).

\subsection{Fraction of blue cluster members}\label{sec:fb}
An examination of the $R$-band image (Fig.~\ref{fig:rbandim}) reveals
that the three clusters are very different in morphological
appearance.  A901a most resembles the traditional picture of a relaxed
cluster dominated by a central bright elliptical galaxy.  A901b also
contains a giant elliptical but the nearby cluster galaxies trace a
more amorphous shape that appears to bleed into the middle of the
supercluster (Fig.~\ref{fig:galaxies}, left).  This is at odds with the
regular, compact X-ray emission detected by \cite{schindler00}.
Finally, A902 is the most irregular of the three, with no well-defined
center.  It is interesting, however, that Fig.~\ref{fig:rbandim}
clearly shows a prominent excess of bright, disky galaxies surrounding
this cluster (absent from the vicinities of the other two).

To better understand the galaxy populations that make up these galaxy
clusters, we calculate the fraction of bright galaxies too blue to
survive the color selection but which may still be cluster members
(presumably late-types).  Taking a 4 arcmin radius aperture, we count
those galaxies with $B-R$ bluer than the cluster sequence of
equation~(\ref{eqn:colcut}) and compare with the total number of galaxies
within the aperture for a given magnitude limit.  We correct these
numbers for background contamination using the number density of
similarly selected galaxies in the complementary `field' region,
taking into account the area obscured by the mask.

\begin{table*}
{\small
\begin{center}
\centerline{\sc Table 3}
\vspace{0.1cm}
\centerline{\sc Fraction of Blue Galaxies for Various Limiting Magnitudes}
\vspace{0.3cm}
\begin{tabular}{rrrrrr}
\hline\hline
\noalign{\smallskip}
\multicolumn{6}{c}{$R<22$}\\
\multicolumn{1}{c}{Cluster} & \multicolumn{1}{c}{$N_{\rm blue,clus}$} & 
\multicolumn{1}{c}{$N_{\rm tot,clus}$} & \multicolumn{1}{c}{$N_{\rm blue,back}$} & 
\multicolumn{1}{c}{$N_{\rm tot,back}$} & \multicolumn{1}{c}{$f_b$}\\
\hline
\noalign{\smallskip}
A901a & 127 & 291 & 82.4 & 158.2 & 0.33\\
A901b &  89 & 234 & 87.5 & 168.0 & 0.02\\
A902  & 137 & 302 & 89.6 & 172.0 & 0.36\\
\hline
\noalign{\smallskip}
\multicolumn{6}{c}{$R<21$}\\
\multicolumn{1}{c}{Cluster} & \multicolumn{1}{c}{$N_{\rm blue,clus}$} & 
\multicolumn{1}{c}{$N_{\rm tot,clus}$} & \multicolumn{1}{c}{$N_{\rm blue,back}$} & 
\multicolumn{1}{c}{$N_{\rm tot,back}$} & \multicolumn{1}{c}{$f_b$}\\
\hline
\noalign{\smallskip}
A901a &  67 & 159 & 38.4 & 74.4 & 0.33\\
A901b & 49 & 120 & 40.8 & 79.0 & 0.20\\ 
A902  & 69 & 154 & 41.7 & 80.9 & 0.37\\
\hline
\noalign{\smallskip}
\multicolumn{6}{c}{$R<20$}\\
\multicolumn{1}{c}{Cluster} & \multicolumn{1}{c}{$N_{\rm blue,clus}$} & 
\multicolumn{1}{c}{$N_{\rm tot,clus}$} & \multicolumn{1}{c}{$N_{\rm blue,back}$} & 
\multicolumn{1}{c}{$N_{\rm tot,back}$} & \multicolumn{1}{c}{$f_b$}\\
\hline
\noalign{\smallskip}
A901a & 33 & 89 & 17.3 & 33.7 & 0.28\\
A901b & 29 & 66 & 18.4 & 35.8 & 0.35\\
A902  & 38 & 81 & 18.8 & 36.6 & 0.43\\
\hline
\noalign{\smallskip}
\multicolumn{6}{c}{$R<19$}\\
\multicolumn{1}{c}{Cluster} & \multicolumn{1}{c}{$N_{\rm blue,clus}$} & 
\multicolumn{1}{c}{$N_{\rm tot,clus}$} & \multicolumn{1}{c}{$N_{\rm blue,back}$} & 
\multicolumn{1}{c}{$N_{\rm tot,back}$} & \multicolumn{1}{c}{$f_b$}\\
\hline
\noalign{\smallskip}
A901a & 15 & 42 & 7.0 & 13.6 & 0.28\\
A901b & 10 & 29 & 7.4 & 14.4 & 0.17\\
A902  & 16 & 29 & 7.6 & 14.8 & 0.58\\
\noalign{\hrule}
\vspace{-1.5cm}
\tablecomments{The final column illustrates the large fraction of bright blue galaxies residing in A902
relative to the other clusters.}
\end{tabular}
\end{center}
}
\label{tab:fb}
\end{table*}

We then calculate
the fraction of bright blue galaxies within the clusters:
\begin{equation}\label{eqn:fblue}
f_b = \frac{N_{\rm blue,cluster} - N_{\rm blue,background}}{N_{\rm total, cluster}
- N_{\rm total,background}}.
\end{equation}
The results tabulated in Table~\ref{tab:fb} confirm that A902 has a
much higher $f_b$ than the other two clusters for a variety of
magnitude limits, ranging as high as $f_b=0.58$ for $R<19$.  We shall
return to the issue of varying cluster luminosities when discussing cluster
$M/L$ ratios in \S\ref{sec:xcorr}.

\section{Weak shear analysis}\label{sec:lensing}

\subsection{Selection of background population}\label{sec-col}
The observed tangential distortion of the background galaxies allows
us to reconstruct the dimensionless surface density of the intervening
matter, $\kappa=\Sigma/\Sigma_{{\rm crit}}$, where
\begin{equation}
\Sigma_{{\rm crit}} = \frac{c^2}{4\pi G D_d}\frac{D_s}{D_{ds}}
\end{equation}
is the critical surface mass density.  To turn the observed $\kappa$
into physical quantities requires knowledge of the angular diameter
distances to the source and lens ($D_d,D_s$), and the relative
distance between the two ($D_{ds}$).  In the case of weak lensing by a
cluster of galaxies, the strength of the lensing signal is governed by
the redshift distribution of the background sources.  This information can
be expressed by the parameter $\beta$, such that
\begin{equation}
\beta = \left< \frac{D_{ds}}{D_s} \right>_{z_s},\  D_{ds} \geq 0.
\end{equation}

\begin{figure}
\centerline{\epsfig{file=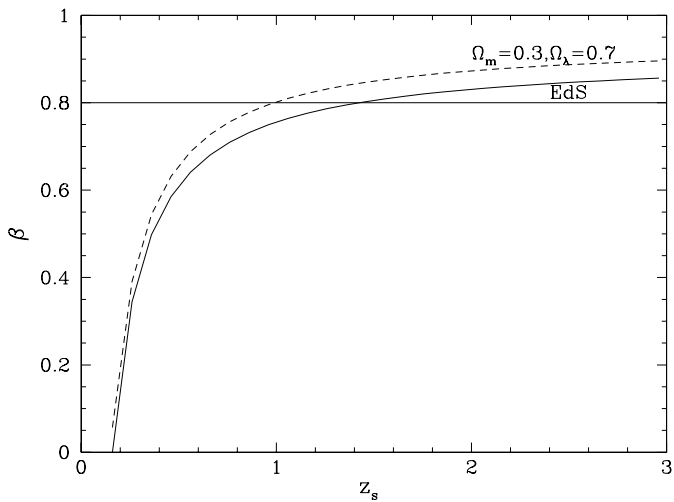,height=0.5\textwidth,angle=270}}
\caption{Dependence of the $\beta$ parameter on the redshift of the
background sources, for a lens at $z=0.16$. Note that the
relatively low redshift of the lens makes it much less sensitive to
the redshift of the background sources for $\langle z_s \rangle \simgt 1$. }
\label{fig:betaplot}
\end{figure}

The galaxies we will use for our mass reconstructions will be composed
for the most part of faint ($R>22$) galaxies or subsets thereof.  It
is of great importance, therefore, that we are able to determine the
appropriate value of $\beta$ to use.  Fig.~\ref{fig:betaplot}a shows
the dependence of the $\beta$ parameter on the redshift of the
background sources, for a lens at $z=0.16$.  Unlike lenses at higher
redshift \citep[see][]{hoekstra00,clowe00}, the relatively
low redshift of our supercluster means that for $z_s \simgreat 1$,
$\beta$ is relatively insensitive to the source redshift, regardless
of the cosmological model used.

The CalTech Faint Galaxy Redshift Survey  \citep[CTFGRS;][]{cohen00} is
a survey in the direction of the Hubble Deep Field North containing
redshifts for 671 galaxies.  To the limits of our observations
($R\sim26$) the median redshift of the CTFGRS implies that the median
redshift of the background galaxies used in our lensing analysis will
be greater than unity.  We therefore adopt the value $\beta=0.8$,
which corresponds to a single screen of background galaxies at
$z_s=1.5$ (for an Einstein-de Sitter cosmology) or $z_s=1.0$ (for
$\Omega_m=0.3,\Omega_\Lambda=0.7$).

\subsection{Tangential shear}\label{sec:gammat}
To measure the weak shear signal from the supercluster, we first
calculate the tangential shear in radial bins around each of the three
clusters, $\gamma_T = -(\gamma_1 \cos 2 \phi + \gamma_2 \sin 2\phi)$.
Here $\phi$ is the azimuthal angle from the center of the mass
distribution, which we take as the optical center (see
Table~\ref{tab:pos}).  To measure the radial surface profiles of the
clusters we use the statistic \citep{fahlman94}:
\begin{equation}
\zeta^{\rm obs}(r,r_{\rm max}) = \frac{2}{1-(r/r_{\rm
max})^2}\int^{r_{\rm max}}_r d\ln(r) \lgl \gamma_T \rgl.
\end{equation}
This provides a lower bound on the surface mass density interior to
radius $r$ relative to the mean surface density in an annulus from $r$
to $r_{\rm max}$:
\begin{equation}
\zeta(r,r_{\rm max}) = \bar\kappa(<r^\prime) - \bar\kappa(r < r^\prime < r_{\rm
max}).
\end{equation}

\begin{figure*}
\centerline{\epsfig{file=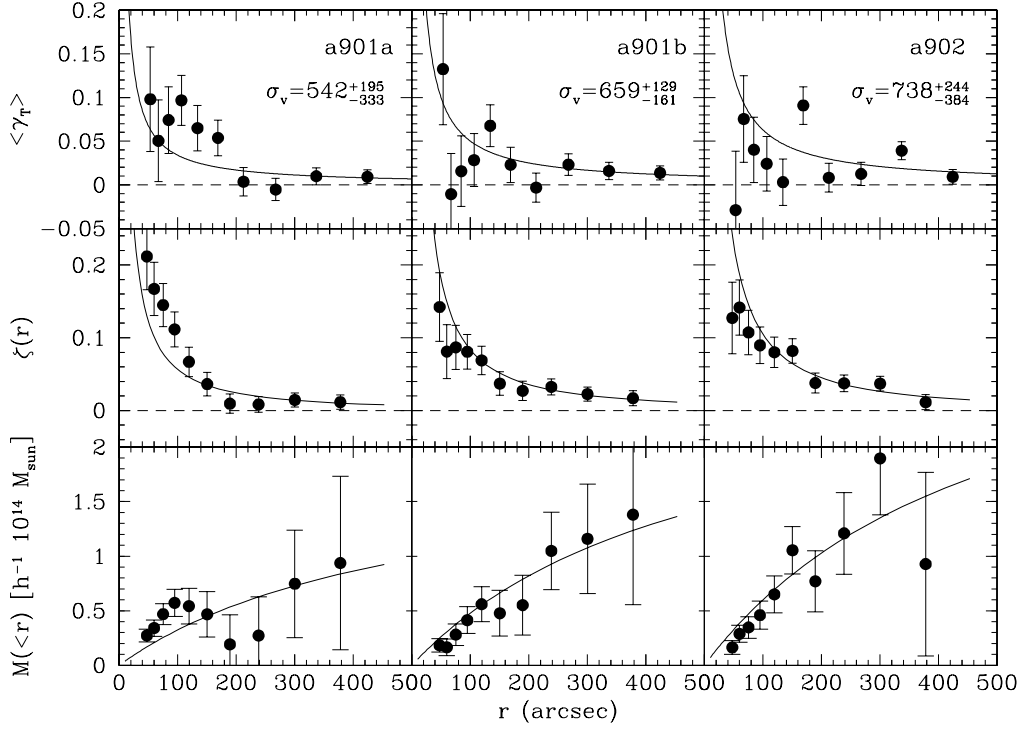,height=0.75\textwidth,angle=270}}
\caption{Radial cluster properties. Top: Tangential shear profiles as
a function of radius from the optical center of clusters A901a (left),
A901b (center), A902 (right). Errorbars are calculated from the
variance in the orthogonal component of the shear \citep[cf.][]{lk97}.
Middle: The corresponding $\zeta$ profile for each cluster, with
uncertainties derived from the uncertainties in the tangential shear.
Bottom: Lower bound on the enclosed mass within a radius $r$ for each
cluster.  In each case the solid line corresponds to the profile
expected for a singular isothermal sphere, fit for each cluster to the
observed tangential shear in radial bins (top row).}
\label{fig:encmass}
\end{figure*}

In Fig.~\ref{fig:encmass} we show the tangential distortion of the
faint ($R>22$) galaxies as a function of distance from the optical
center for each of the three clusters: A901a, A901b, and A902.
Non-zero tangential shear is observed in each case out to a
significant radius.

To model the mass of each cluster, we fit a singular isothermal sphere
to the observed tangential distortions.  The best-fit models yield
velocity dispersions $\sigma_v = 542^{+195}_{-333}$,
$659^{+129}_{-161}$, and $738^{+244}_{-384}$ km s$^{-1}$ for A901a,
A901b, and A902 respectively.  The fits to the tangential shear
profiles are shown in the top panels of Fig.~\ref{fig:encmass}.  Also
shown in the central panels of Fig.~\ref{fig:encmass} is $\zeta(r)$
for each cluster (note that the points are not independent as each
depends on the mass profile of the cluster exterior to $r$).  A value
of $r_{\rm max}=500$ arcsec was used.

The A901a cluster, while shown by \cite{schindler00} to be not as X-ray
luminous as A901b is certainly the most regular `looking' cluster of
the three.  It appears to have a regular and relaxed morphology with a
prominent central bright elliptical galaxy.  However, the tangential shear
profile is significantly inconsistent with that expected for an
isothermal sphere.  The strong shear signal measured in the inner
annuli ($80\arcsec<r<200\arcsec$) drops dramatically at larger radii, resulting in a
$\zeta$ profile considerably steeper than isothermal.  The sudden drop
in signal could perhaps be explained by an encroachment into the
neighboring cluster A901b (although the profile for that cluster
shows no similar truncation).

The integral of the $\zeta$ statistic gives a lower bound on the
mass enclosed within a radius $r$, $M(<r)=\pi r^2\zeta(r)\Sigma_{\rm
crit}$, which is shown in the bottom panels of Fig.~\ref{fig:encmass}.
Again, the solid lines correspond to the profile expected from the
singular isothermal sphere fitted to the tangential shear.  

\subsubsection{Comparison with X-ray analysis}
At this point it is constructive to compare the results of the
tangential shear analysis with that of the previous X-ray studies.  With a
pointed HRI image of the A901/902 field, Schindler resolved several
point sources in the region that contributed to the original X-ray
fluxes of  \cite{ebeling96}.  The X-ray emission from A901a was
found to be consistent with a point source, and A901b showed compact
cluster emission with a revised $f_X(0.1-2.4) {\rm keV} = 3.0\times
10^{-12}$ erg s$^{-1}$ cm$^{-2}$.

Schindler estimates $T_X=4$ keV from $L_X - T$ relations for A901b.
We use this temperature, together with the empirically determined
relations of \cite{girardi96} to compute
\begin{equation}
\sigma_v = 10^{2.53\pm 0.04}T^{0.61\pm 0.05}.
\end{equation}
This yields $\sigma_v\simeq 800$ km s$^{-1}$, which is slightly higher
than the estimate derived from the tangential shear analysis above but
consistent within the 1$\sigma$ error bound.  The discrepancy is not
surprising, however, as the estimates assume spherical symmetry and
require the cluster to be in hydrostatic equilibrium.  Given the close
proximity of A901a and evidence for interaction between the two
clusters seen in the distribution of galaxies, this assumption is not
likely to be a valid one.

Finally, we reexamine the location of the Schindler X-ray sources in
light of our deep $R$-band image.  Fig.~\ref{fig:rbandim} shows that
the compact X-ray emission of Abell 901b originates from the location
of the central bright elliptical galaxy (which we have adopted as the
optical center of this cluster).  The downgraded X-ray emission of
A901a is coincident with a bright cluster galaxy several arcseconds
east of the central elliptical in that cluster.  In both cases the
NRAO/VLA All Sky Survey \citep{condon98} shows that these are strong
radio sources, with flux equal to 93.9 mJy (A901a) and 20.5 mJy
(A901b), supporting the conclusion of \cite{schindler00} that some of
the X-ray emission is due to AGN activity.  Lastly, considering the
X-ray emission near A902, it is clearly not related to the cluster
core but could be associated with one of several faint $R$-band
sources nearby.

In short, while the mass derived from the X-ray emission from A901b is
in broad agreement with our estimate from the weak lensing analysis,
the X-ray data on the whole paint a markedly different picture than
that revealed by weak lensing.  Two of the three clusters (including
the most massive, A902) are undetected in X-rays.  In the next section
we shall see how the supercluster structure and dark matter
distribution is revealed by non-parametric weak lensing reconstructions.

\begin{table*}
{\small
\begin{center}
\centerline{\sc Table 4}
\vspace{0.1cm}
\centerline{\sc Cluster Positions and Parameters Derived from the Weak
Lensing Analysis}
\vspace{0.3cm}
\begin{tabular}{rrrrcrrrcr@{\,$\pm$\,}l}
\hline\hline
\noalign{\smallskip}
\multicolumn{1}{c}{Cluster} & 
\multicolumn{1}{c}{$\alpha$} &
\multicolumn{1}{c}{$\delta$} & 
\multicolumn{1}{c}{$\sigma_v$} & 
\multicolumn{1}{c}{Aperture} & 
\multicolumn{1}{c}{$\bar{\kappa}_M(<r)$} & 
\multicolumn{1}{c}{$\nu$} & 
\multicolumn{1}{c}{$M(<r)$}  & 
\multicolumn{1}{c}{$L(<r)$} &
\multicolumn{2}{c}{$M/L$} \\
 &
\multicolumn{1}{c}{\scriptsize (J2000)} &
\multicolumn{1}{c}{\scriptsize (J2000)} &
\multicolumn{1}{c}{\scriptsize (km s$^{-1}$)} &
 & & & 
\multicolumn{1}{c}{\scriptsize ($h^{-1} 10^{13}{\rm M_{\sun}}$)} &
\multicolumn{1}{c}{\scriptsize ($h^{-2} 10^{11}L_{B_{\sun}}$)} & 
\multicolumn{2}{c}{\scriptsize ($h {\rm M_{\sun}/L_{B_{\sun}}}$)}\\
\hline
\noalign{\smallskip}
A901a &  09:56:26.4 &  -09:57:21.7 & $542^{+195}_{-333}$ &
$r=1\arcmin$ & $0.093\pm 0.028$ & 3.2 & $1.95\pm0.59$  &
1.85 & $104$&$32$ \\ 
 &  &  &  & $r=2\arcmin$ & $0.046 \pm 0.022$ & 2.0 &
 $3.78\pm 1.83$ & 2.41 & $156$&$75$\\ 
 &  &  &  & $r=3\arcmin$ & $0.015\pm 0.017$ & 0.8 & 
 $2.76\pm 3.45$ & 3.63 & $ 75$&$ 94$\\
\\

A901b &  09:55:57.4 & -09:59:02.7   & $659^{+129}_{-161}$ &
$r=1\arcmin$ & $0.108\pm0.024$ & 4.4 & $2.27\pm
0.51$ & 0.62 & $363$&$ 82$\\
 & &  &  & $r=2\arcmin$ & $01.079\pm0.019$ & 4.1 &
$6.50\pm 1.58$ & 0.85 & $764$&$ 186$\\ 
 &  &  &  & $r=3\arcmin$ & $0.053\pm 0.015$ & 3.6 &
$9.83\pm 2.37$ & 2.36 & $416$&$ 116$\\
\\

A902  &  09:56:33.6 & -10:09:13.1 & $738^{+244}_{-384}$ &
$r=1\arcmin$ & $0.115\pm 0.024$ & 4.7  & $2.36\pm
0.45$ & 1.16 & $204$&$ 43$\\
 &  &   &   & $r=2\arcmin$ & $0.084\pm 0.021$ & 4.0 &
$6.85\pm 1.70$ & 1.92 & $356$&$ 89$ \\
 &  &   &   & $r=3\arcmin$ & $0.053\pm 0.016$ & 3.2 &
$ 9.75\pm 3.05$ & 2.20 & $442$&$ 138$ \\
\noalign{\hrule}
\vspace{-1.5cm}
\tablecomments{Columns 2 and 3 give the coordinates of the optical
centres of the clusters used in this analysis.  Column 4 gives
the velocity dispersion of the best-fitting SIS model (see
\S\ref{sec:gammat}) derived from the parameterized fit to the
tangential shear around each cluster.  The remaining columns list data
for the two dimensional mass reconstructions of \S\ref{sec:2d} and the
$M/L$ ratios derived in \S\ref{sec:ml} in apertures of radius $r$
(Column 5).}
\end{tabular}
\end{center}
}
\label{tab:pos}
\end{table*}

\subsection{Non-parametric mass reconstructions}\label{sec:2d}

In addition to the parametrised fits to the individual clusters
described above, we also used the parameter-free mass reconstruction
algorithm of \cite{ks93} to construct two-dimensional
maps of the surface mass density.  As with the color-selected
foreground population described above, we use different
magnitude/color selection criteria to isolate samples of `red',
`blue', and `faint' background galaxies for use in the lensing
reconstruction.  The properties of these populations are summarized in
Table~\ref{tab:pops}.

\begin{table*}[t]
{\small
\begin{center}
\centerline{\sc Table 5}
\vspace{0.1cm}
\centerline{\sc Selection Criteria Defining the Galaxy Populations Used in
the Weak Lensing Mass Reconstructions}
\vspace{0.3cm}
\begin{tabular}{cccc}
\hline\hline
\noalign{\smallskip}
Population & Selection Criteria & $N$ & $n$(arcmin$^{-2}$)\\
\hline
\noalign{\smallskip}
red	&	$B-R >$ cluster	&  22296   & 21.1 \\
blue    &	$B-R <$ cluster	&  18736   & 17.7 \\
faint   &       $R>22$		&  40485   & 38.3 \\
\noalign{\hrule}
\vspace{-1.5cm}
\tablecomments{The `blue' and `red'
populations are those galaxies that posses $B-R$ colors less than
and greater than the cluster color-magnitude sequence shown in
Fig.~\ref{fig:colmagfieldclus}, respectively.  The third and fourth
columns list the total number and surface number density of galaxies
for each population.  Note that the blue-selected galaxies are more
likely to contain foreground objects and blue cluster galaxies.}
\end{tabular}
\end{center}
}
\label{tab:pops}
\end{table*}

\begin{figure*}
\centerline{\epsfig{file=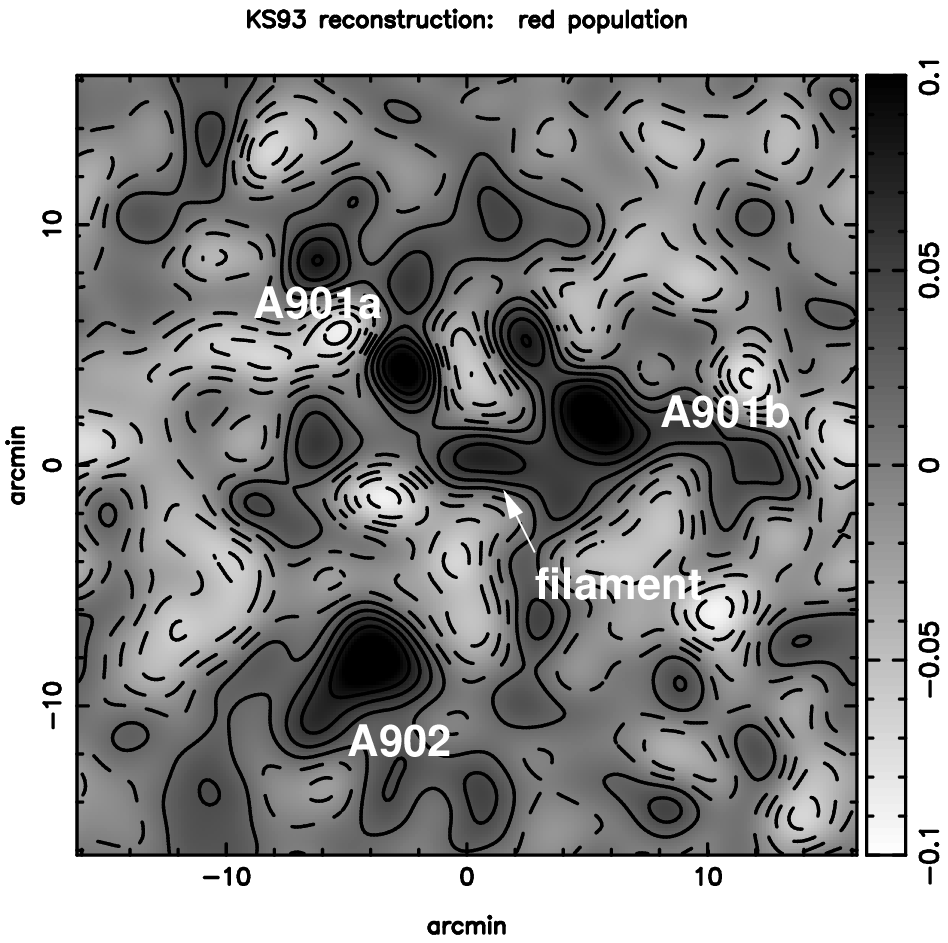,width=0.5\textwidth,angle=0}
            \epsfig{file=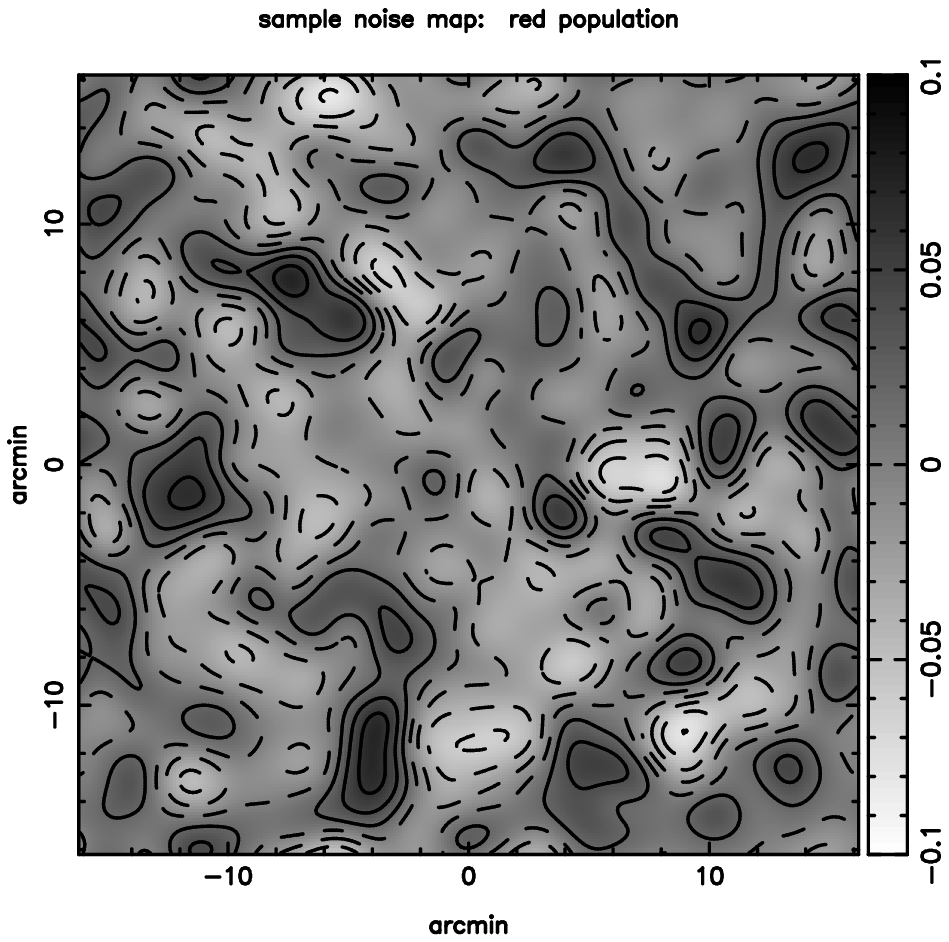,width=0.5\textwidth,angle=0}}
\centerline{\epsfig{file=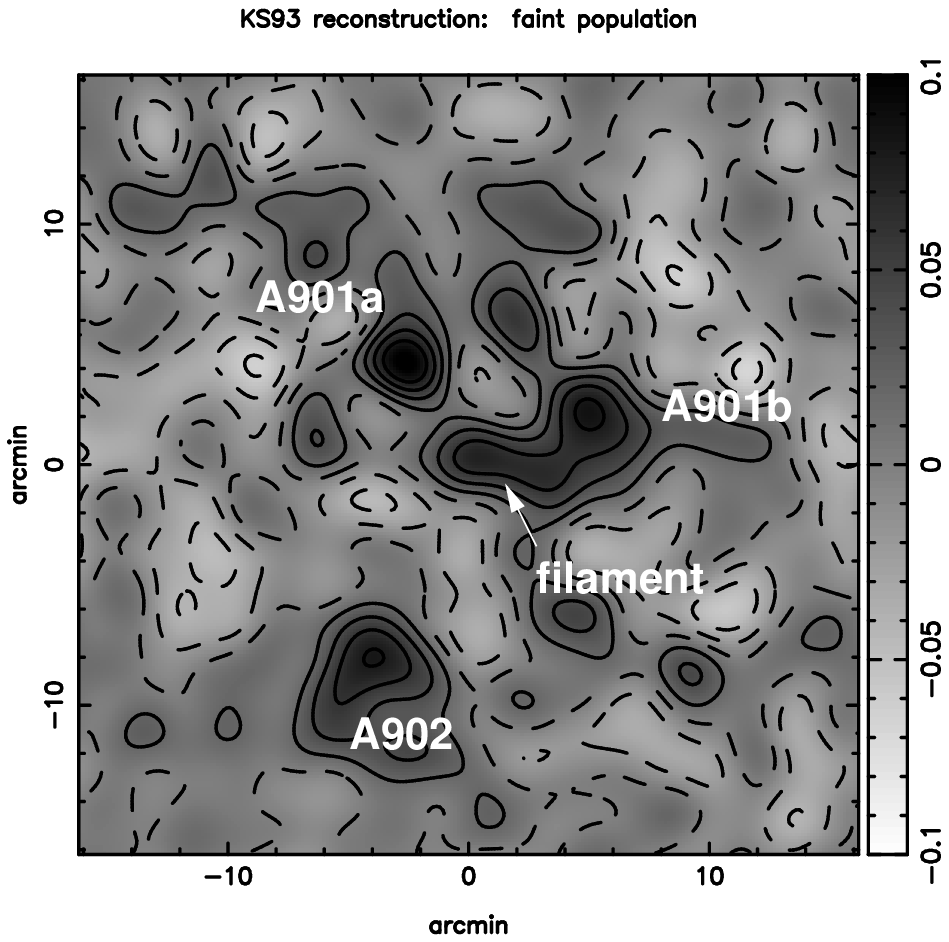,width=0.5\textwidth,angle=0}
            \epsfig{file=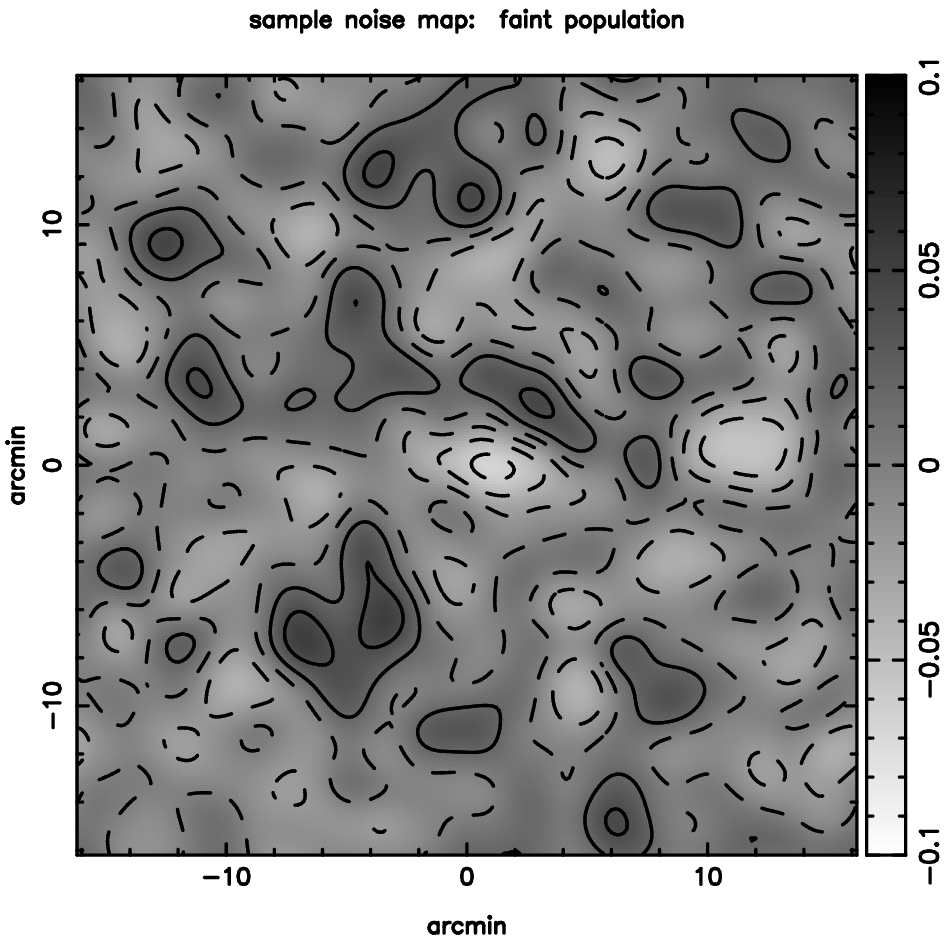,width=0.5\textwidth,angle=0}}
\caption{Non-parametric mass reconstructions and noise levels. Left:
Dimensionless surface mass density, $\kappa_M$, reconstructions
applying the \cite{ks93} algorithm to the `red' (top) and `faint'
(bottom) background populations (not independent).  All three clusters
appear as prominent mass peaks, and there is evidence in both maps of
possible filamentary structure between A901b and A901a.
Right: Demonstration of the noise levels for each mass map,
created by randomly shuffling the ellipticities of the background
galaxies but maintaining the same positions prior to the
reconstruction.  All reconstructions are smoothed with a $\sigma=60$
arcsec Gaussian.}
\label{fig:redfaintmass}
\end{figure*}

It is immediately apparent from examining the mass reconstructions
obtained from the `red' and `faint' samples of background galaxies
(shown in Fig.~\ref{fig:redfaintmass}) that the mass peaks
corresponding to the three clusters are each strongly detected and are
fairly regular in appearance.  The magnitude of the mass concentrations
follows the same ordering according to mass as inferred from the
strength of the tangential shear profiles (from lowest to highest
mass: A901a, A901b, A902).  In addition, there is evidence of
filamentary structure connecting A901a and A901b.  We shall return to
this in more detail in \S\ref{sec:filament}.

To gain an estimate of the noise levels associated with these mass
reconstructions, we randomly shuffled the ellipticities of the background
galaxies while maintaining their positions, and performed the mass
reconstruction for each of 32 realizations of the background galaxies.
Sample noise maps for the two reconstructions are shown in the right
panels of Fig.~\ref{fig:redfaintmass}.  These levels are in agreement
with the variance predicted from the KS93 algorithm, given the
smoothing scale $\theta$ and mean surface density $\bar{n}$:
\begin{equation}\label{eqn:kappavar}
\lgl\kappa^2 \rgl = \frac{\lgl\gamma^2\rgl}{8\pi \bar{n} \theta^2 }.
\end{equation}
Assuming a value of $\lgl\gamma^2\rgl\simeq0.2$ for the variance
due to the intrinsic shapes of galaxies, equation~(\ref{eqn:kappavar})
predicts $1\sigma$ uncertainty in $\kappa$ of 0.019 for the `red' mass
map and 0.014 for the `faint' mass map.

The reconstruction obtained from the `blue' sample of galaxies is
markedly different from the red reconstruction, however.  In the blue
map (not shown), the cluster peaks are not present or are offset from their
positions in the red map.  In addition, there is an additional
prominent mass peak west of A902 that does not appear in the red map,
nor does it appear to be associated with any concentration of bright
foreground galaxies.  The dilution of the lensing signal from the
clusters is not entirely surprising, given that by selecting galaxies
with $B-R$ colors bluer than the supercluster galaxies one would
expect to select mostly foreground galaxies or blue cluster members.

Furthermore, progressive slices in $R$-band magnitude of the blue
catalogue reveal that the new mass peak is due to the ellipticities of
only the faintest $(R>26)$ galaxies, for which the shape measurements
are least reliable. We therefore conclude that the peak in the mass map
derived from the `blue' sample is likely to be spurious and the result
of noise.  In fact, this confirms that most of the lensing signal seen
in the faint mass reconstruction is contained in the redder galaxies,
and that the noise estimate for the faint reconstruction has been
somewhat underestimated as the actual number density $n$ of galaxies
contributing to the lensing signal is less than quoted in
Table~\ref{tab:pops}.

\subsection{Detection of a dark matter filament}\label{sec:filament}

The mass reconstructions of Fig.~\ref{fig:redfaintmass} show evidence
of an extension of the mass distribution of A901b in the direction of
A901a.  This is supported by the presence of an elongation of the
distribution of galaxies in that region (Fig.~\ref{fig:galaxies}).
However, we note that the most prominent part of this mass `filament'
extends in the east-west direction and is located almost in the center
of our image.  This is very close to the intersection of the corners
of four of the subregions on which we performed the PSF correction and
so is worthy of reexamination.  While we saw no sharp discontinuities
in the behavior of the PSF at this location, the stars used for the
corrections do only finitely sample the PSF on the chip and the bicubic
polynomials could fail to properly correct this region if the
anisotropies were large and varying.

To test the robustness of the PSF correction and the recovery of the
filament in the mass maps, we first redid the PSF correction by
fitting a seventh-order two-dimensional polynomial across the entire
field (rather than applying our previous method of applying a
lower-order correction separately to each subregion).
However, this higher-order polynomial correction still left the
largest residuals in the corrected stellar ellipticities in the
central regions of the image.

Since shear is a non-local measurement, and considering that the
catalogue of background galaxies already contains a small number of
regions masked out due to contamination by bright stars, we then
tested the effect of removing several regions of the catalogue in
which the ellipticity measurements might be less reliable.  Excising
the central $2\arcmin\times 2\arcmin$ from the catalogue of background
galaxies, we again performed the mass reconstructions.  We found that
the peak of the filamentary structure fell by $\Delta\kappa\simeq0.02$
but that the underlying plateau connecting A901a and A901b was still
present at the $\kappa\simeq 0.04-0.06$ levels (i.e., $\sim 2-3\sigma$
above the noise).  While having holes in the input catalogues is not
ideal for the purposes of the mass reconstruction, this gives some
indication that at least some of the signal is real and significant.
Similar results were found when we removed not only the central
regions but also masked out the intersections of the regions used in
the PSF correction and the regions in the corners of the image.  These
are regions in which any departures from the linear astrometric
corrections of \S\ref{sec:astrometry} could cause objects to be
misaligned and would lead to a smearing of their shapes in the final
combined image, which could then be misinterpreted as a lensing
induced shear.  However, even after such conservative cuts to the
object catalogue, the filamentary structure connecting A901a and A901b
continued to be detected in the mass reconstructions at a level
significantly above the noise.

As an additional test, we attempted to perform the mass-reconstructions on
object shape parameters derived from the $B$-band image, to check if
the filament appeared in reconstructions in both bands.  However, the
less favorable atmospheric conditions and shorter exposure time for
this image meant that we were unable to measure reliable $B$-band
shapes for a sufficient number of galaxies to create a mass map.

The fact that the observed filament does not lie directly along the
axis connecting the structure, but instead arcs from one to the next
could be reflective of the initial orientation of the clusters.
\cite{bond96} demonstrate how the tidal gravitational field at
location of initial density peaks can create filaments with a variety
of orientations and density contrasts, including the qualitative shape
observed here.  Given that matter filaments have long been a robust
prediction of the dark matter scenario, regardless of the nature of
the dark matter, this detection provides good confirmation of their
existence, and hence generally supports the dark matter scenario.

\section{Comparing mass and light}\label{sec:xcorr}

\begin{figure*}
\centerline{\epsfig{file=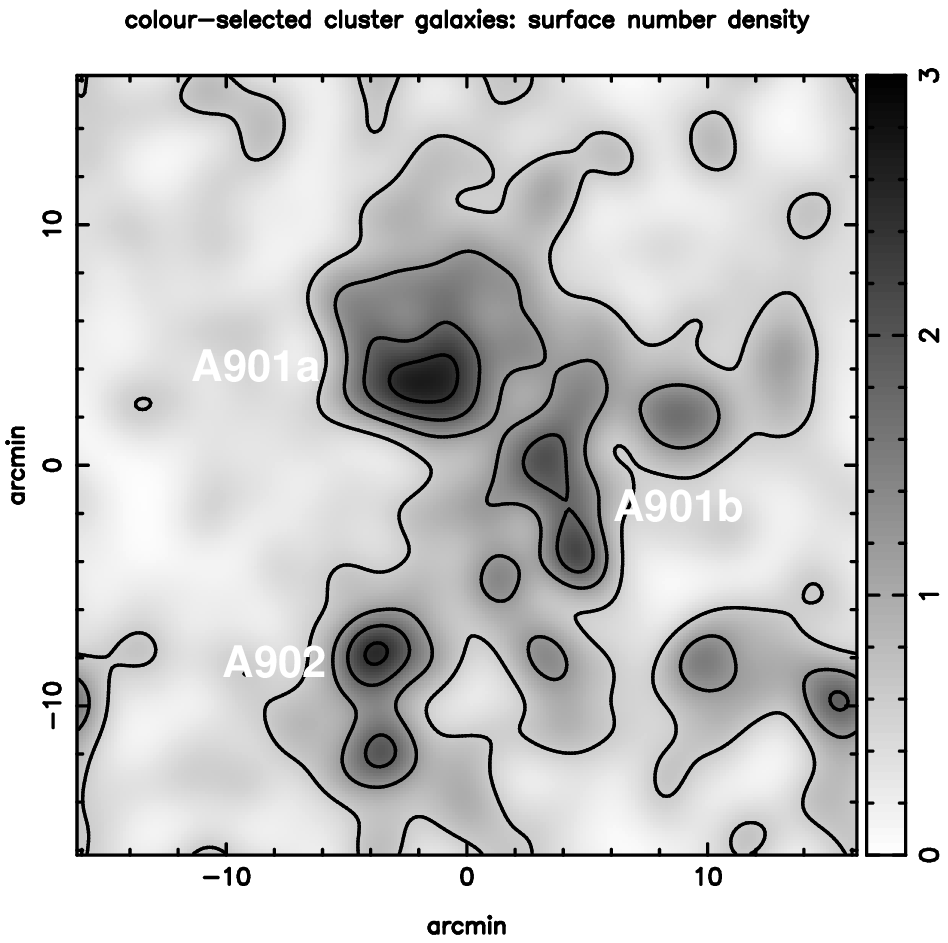,width=0.5\textwidth,angle=0}
            \epsfig{file=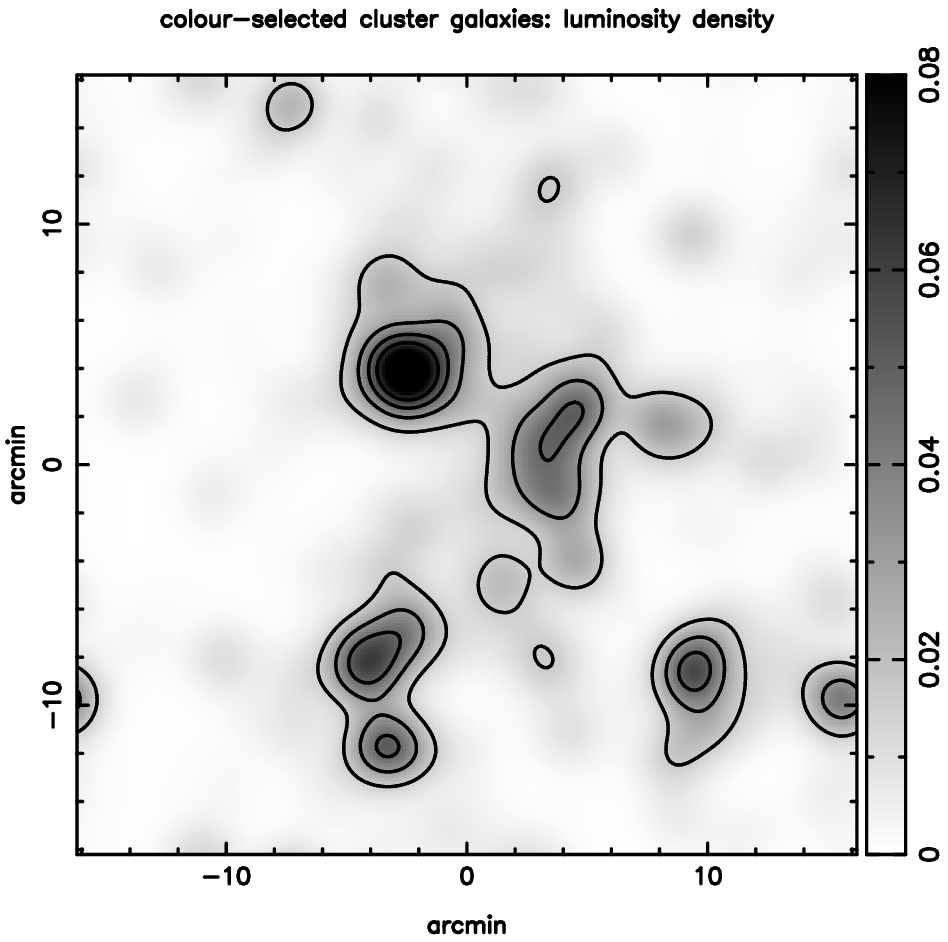,width=0.5\textwidth,angle=0}}
\caption{Smoothed distributions of supercluster galaxy density and
luminosity. Left: Surface number density of the bright ($R<20$)
color-selected cluster galaxies (arcmin$^{-2}$), smoothed with a
$\sigma=60$ arcsec Gaussian.  Right: Luminosity-weighted distribution of the
same galaxies, showing the prominence of the A901a light
distribution.}
\label{fig:lumndens}
\end{figure*}

Comparison of the mass reconstructions with the smoothed surface
number density and luminosity maps presented in
Fig.~\ref{fig:lumndens} reveals two different descriptions
of the supercluster.  Contrary to the mass maps, in both light and
number density the most prominent of the three clusters is by far
A901a.  Furthermore, the optical appearance of A901b is much more
elongated and much less regular than the mass reconstructions (and the
compact nature of the X-ray emission from this cluster) would suggest.

\subsection{$M/L$ ratios in apertures}\label{sec:ml}
We estimate actual $M/L$ ratios for each cluster by computing the ratio of
the total mass inferred from the two-dimensional reconstructions with the
total $B$-band luminosity of the color-selected early-type galaxies
within an aperture.  Note that the mass is subject to a possible
upward correction due to the insensitivity of the weak shear method to
a uniform mass distribution along the line of sight (the `mass-sheet
degeneracy'), so formally our mass estimates, and hence $M/L$ ratios,
are lower limits.  However, the field of view of our image is
sufficiently large (and the clusters themselves are well-removed from
the edges) that we may assume that $\kappa\rightarrow 0$ at the edge of
the reconstructions.

The $M/L$ values listed in Table~\ref{tab:pos} show that the clusters
display a range of $M/L$ ratios.  We find that the $M/L$ ratios for
A901a are significantly lower for all aperture sizes than the values
for A901b and A902, which are in much better agreement with each
other.  We note, however, that the $M/L_B$ ratios quoted here consider
the total mass, but only the contribution of the {\it early-type} galaxies
to the luminosity.  The large fraction of bright blue galaxies seen in
Abell 902 could serve to increase the total luminosity of the cluster
and thus lower the $M/L$ ratio for that cluster.  The discrepancies
between the values for all three clusters would, however, still
remain.

\begin{figure}
\centerline{\epsfig{file=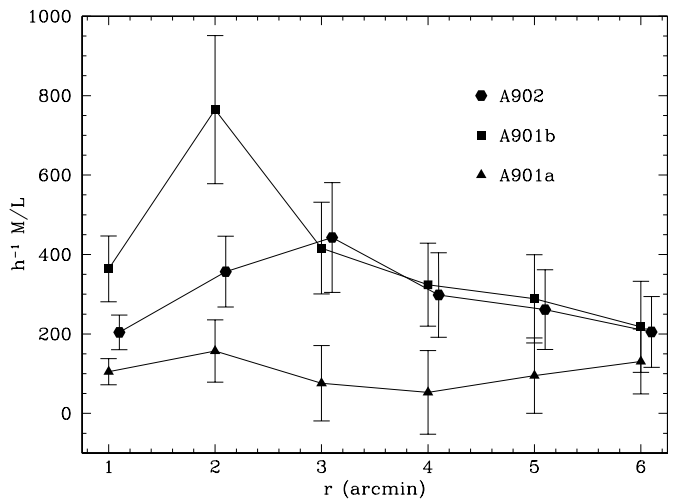,height=0.5\textwidth,angle=270}}
\caption{$M/L$ ratios with increasing aperture size for each of the
three clusters (points for A902 have been slightly shifted
horizontally for clarity).  There is no strong evidence for a change
in $M/L$ ratio with scale for any of the clusters, except for the
spike at $r=2$ arcmin for A901b which reflects the misalignment of
mass and light in this cluster.  A901a shows a significantly lower $M/L$
ratio than the other clusters for all aperture sizes.  The apertures
for A901a and A901b begin to overlap at $r=3.9$ arcmin.}
\label{fig:ml}
\end{figure}

Fig.~\ref{fig:ml} shows the resulting $M/L$ ratios for each cluster as
the aperture size is varied.  There is no evidence for an increase of
$M/L$ ratio with aperture size for any of the clusters beyond 2 arcmin
($0.2h^{-1}$ Mpc), implying that there is no appreciable reservoir of
dark matter in the inter-cluster regions.  The values appear to
converge at $M/L\sim 200h$ at large radii, albeit with some overlap
between apertures.  The dark matter distribution appears no more
extended than the distribution of early-type galaxies (although as we
have seen previously in the case of A901b, the two are not always
precisely co-located).

\subsection{Predicting mass from light}\label{sec:mass-light}
So that we may further compare mass and light across the whole field
of view, we next create a prediction of the surface mass density,
$\kappa_L$, from the luminosity of the color-selected supercluster
galaxies.  We determine the $k$-correction appropriate for early-type
galaxies at $z=0.16$ using the population synthesis models of
\cite{bc96} for a passively evolving early-type galaxy formed
with a single burst of star formation at $z=5$.  We convert observed
$B$-band magnitude into luminosity, $L_B$, for each foreground galaxy.
We can then calculate the the contribution of each galaxy to the
surface mass density in a region by $mass\ density = light \times
(M/L) / area$.  Dividing by the critical surface mass density we find
the dimensionless surface mass density
\begin{equation}\label{eqn:kappal}
\kappa_L d\Omega = \frac{L_B 
M/L}{D_d^2}\frac{1}{\Sigma_{\rm crit}} = \frac{L_B
300h\Upsilon M_\odot/L_\odot}{D_d^2 \Sigma_{\rm crit}}.
\end{equation}
As a first estimate, we assume a constant $M/L_B = 300h$, $\Upsilon = 1$
to create the predicted $\kappa_L$ map.  This will be revised later by
varying the normalization $\Upsilon$.  We choose to predict the mass
from the observed light, rather than the other way round, because
formally the uncertainties are less:  we can observe the positions and
luminosities of the galaxies to create the light map.

The resulting $\kappa_L$ differs from the mass map recovered from our
weak shear analysis in that it is non-negative everywhere with
different noise properties.  So that we are able to directly compare
$\kappa_L$ and $\kappa_M$, we follow the procedure of \cite{kaiser98}
 and \citet*[][hereafter WLK]{wilson01}.  Using the estimate for
$\kappa_L$ constructed as described in equation~(\ref{eqn:kappal}) but
smoothed on a finer scale, we solve the two-dimensional Poisson
equation to recover the projected lensing potential.  We then derive a
shear field from this potential and sample the field at the location
of our background galaxy catalogue to emulate the finite sampling of
our observations.

\begin{figure}
\centerline{\epsfig{file=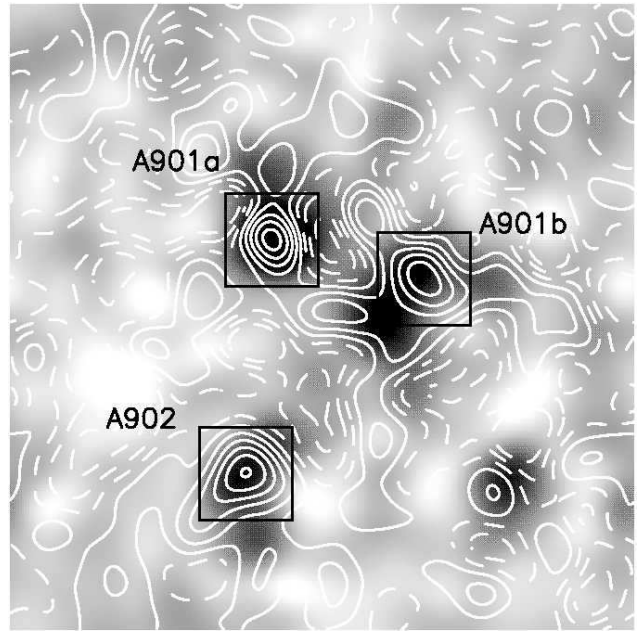,width=0.45\textwidth}}
\caption{Comparison of surface mass density $\kappa$ predicted from
the luminosity of early-type galaxies ($\kappa_L$, gray-scale) and that
recovered from the weak shear reconstructions ($\kappa_M$, contours).
The light from the early-type galaxies generally traces the mass,
except in the case of A901b.  Note that the clump of supercluster
galaxies west of A902 (see Fig.~\ref{fig:galaxies} shows up as a
prominent fourth peak in the $\kappa_L$ map.  The mass contours show a
slight overdensity in $\kappa_M$ near this location, but not
necessarily one that would be identified as a structure from the
lensing signal alone.}
\label{fig:lightmasscont}
\end{figure}

At this point the predicted shear field is still idealized, as it does
not contain the noise associated with the intrinsic ellipticity of the
background sources.  To each value sampled from the shear field, we
therefore add a random noise component drawn from a Gaussian
distribution with $\sigma=0.3$, which reflects the measured noise from
our galaxy catalogue.  We then apply the KS93 algorithm to the
predicted catalogue, to produce a new $\kappa_L$ with the same
finite-field effects and similar noise properties as our measured
$\kappa_M$ (although still insensitive to any structure outside the
field of view).

The resulting predicted surface mass density, $\kappa_L$, and that
recovered from the weak shear reconstructions, $\kappa_M$, are
overlayed in Fig.~\ref{fig:lightmasscont}.  The light from the early
type galaxies traces the location of the mass fairly well, with the
notable exception of the elongated optical appearance of A901b, which
is also displaced from the associated strongly detected mass peak.  A
fourth peak is predicted in the $\kappa_L$ map at the location of the
subclump of supercluster galaxies west of A902 (see
Fig.~\ref{fig:galaxies}, right), and this is marginally reproduced in the
$\kappa_M$ map.

\begin{figure*}
\centerline{\epsfig{file=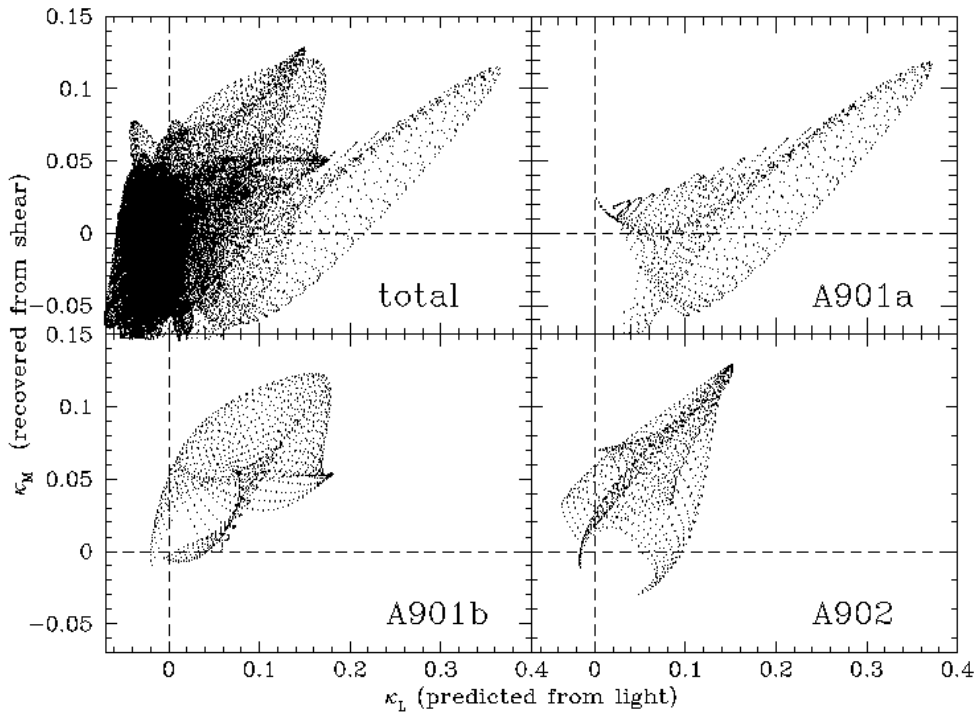,height=0.75\textwidth,angle=270}}
\caption{Pixel-by-pixel comparison of observed $\kappa_M$ and
predicted $\kappa_L$ maps.  The top left panel represents the entire
field of view, while the remaining panels display pixels within the boxes
around each of the clusters shown in Fig.~\ref{fig:lightmasscont}.
Note that the misalignment of light and mass in A901b produces a much
broader distribution of points than the more linear relations for
A901a and A902.}
\label{fig:scatterboxplot}
\end{figure*}

A pixel-by-pixel comparison of the two $\kappa$ maps is shown in
Fig.~\ref{fig:scatterboxplot}.  Additionally, the scatterplots
for the regions centered around each mass peak (indicated by the boxes
in Fig.~\ref{fig:lightmasscont}) are shown separately for
clarity.  The broad distribution of the A901b points relative to the
other two illustrates the general misalignment between light and mass
in this cluster.  Conversely, the close alignment between mass and
light for the remaining two clusters is reflected in the much narrower
relation.  The slope of the linear relation in each case reflects the
degree to which the $M/L$ ratio assumed for the predicted $\kappa_L$ map
approximates the true value for that cluster.  The differing slopes
for A901a and A902 confirms that a single $M/L$ does not satisfy both
clusters.

\subsection{Cross-correlation of mass and light}

\subsubsection{Global cross-correlations}
While in \S\ref{sec:ml} and \S\ref{sec:mass-light} we investigated the
local properties of the mass and light, it is also useful to study the
\textit{global} properties of the field.  In this section, we will
perform a statistical cross-correlation and auto-correlations of the
predicted $\kappa_L$ and measured $\kappa_M$ mass maps.  We shall
assume a simple linear biasing model, in which the light is related to
the mass by a constant, scale-independent $M/L$ ratio.  We recognize
that this is unlikely to be the optimal model given the scatter in the
cluster $M/L$ ratios (cf. Fig.~\ref{fig:ml}).  However, such a model
will serve as a relatively simple starting point for this analysis and
has the added advantage of allowing us to compare results with those
of \cite{kaiser98}.  We write
\begin{equation}\label{eqn:bias}
\kappa_L = \Upsilon^{-1}\kappa_M + \epsilon,
\end{equation} 
where $\epsilon$ is a stochastic component \citep{dekel99} with
variance $\sigma_\epsilon^2$, and $\Upsilon$ is the normalization
required to correct the assumed $M/L = 300h$ of equation~(\ref{eqn:kappal})
to the true $M/L$ ratio.  The stochastic component, $\epsilon$,
reflects the hidden variables due to all the nonlinear, nonlocal
influences on galaxy and star formation not directly associated with
the local density field.  Note that $\lgl \epsilon \rgl = 0$.  Later
we will discuss the possibility that $\Upsilon$ is some more
complicated function of scale or mass, but for the following analysis
make the assumption of \cite{kaiser98} that a single, constant,
scale-independent $M/L$ ratio can describe the whole supercluster
region.

Although there are physical arguments for starting with mass and
attempting to discover a relation that allows one to recover the
light, for practical reasons we shall follow the opposite route and
consider
\begin{equation}\label{eqn:bias2}
\kappa_M = \Upsilon(\kappa_L - \epsilon).
\end{equation}
As discussed earlier, this is motivated by the fact that our
measurement of the mass is relatively noisy compared to the light.

We define the two-dimensional cross-correlation of two images $A$ and
$B$ as 
\begin{equation}\label{eqn:xcorrdef} 
C^{AB}({\delta\theta}) = \lgl\kappa_A({
\theta})\kappa_B({\theta+\delta\theta})\rgl
\end{equation}
where we average over the products of all pairs of pixels separated by
$\delta\theta$.

Using the linear biasing model of equation~(\ref{eqn:bias2}), the cross-correlation of
$\kappa_M$ and $\kappa_L$ becomes
\begin{equation}\label{eqn:cml}
\begin{array}{rcl}
C^{ML} & = & \lgl \kappa_L \kappa_M \rgl \\
       & = & \lgl \kappa_L[\Upsilon(\kappa_L-\epsilon)]\rgl\\
       & = & \Upsilon\lgl\kappa_L^2\rgl,
\end{array}
\end{equation}
since $\epsilon$ and $\kappa_L$ are assumed to be uncorrelated.
Similarly, the autocorrelation of $\kappa_M$ becomes
\begin{equation}\label{eqn:cmm}
\begin{array}{rcl}
C^{MM} &  = & \lgl \kappa_M \kappa_M \rgl\\
       &  = & \lgl([\Upsilon(\kappa_L - \epsilon)][\Upsilon(\kappa_L - \epsilon)]\rgl\\
       &  = & \Upsilon^2\lgl\kappa_L^2\rgl + \Upsilon^2\sigma_\epsilon^2,
\end{array}
\end{equation}
and the autocorrelation of $\kappa_L$ is simply
\begin{equation}\label{eqn:cll}
C^{LL} = \lgl\kappa_L^2\rgl.
\end{equation}
We can therefore calculate all three of these relations and determine
values for $\Upsilon$ and $\sigma_\epsilon^2$.

However, to obtain an unbiased estimate of these correlation
functions, we need to remove the noise, as in 
\begin{eqnarray}
C^{LL}_{\rm true} & = & C^{LL}_{\rm obs} - C^{LL}_{\rm noise},
\label{eqn:clltrue} \\ 
C^{MM}_{\rm true} & = & C^{MM}_{\rm obs} - C^{MM}_{\rm noise},
\label{eqn:cmmtrue}\\ 
C^{ML}_{\rm true} & = & C^{ML}_{\rm obs}. \label{eqn:cmltrue}
\end{eqnarray}
Note that the noise properties of $\kappa_L$ and $\kappa_M$ are uncorrelated.

We estimate $C^{MM}_{\rm noise}$ by constructing an ensemble of 32
mass maps from randomized versions of our faint galaxy catalogues in
the same manner and with the same smoothing scale as $\kappa_M$.  For
each realization, the positions of the background galaxies remain
fixed while the ellipticities of the galaxies are shuffled, and these
noise maps (e.g. the right-hand panels of Fig.~\ref{fig:redfaintmass})
are used to calculate the average autocorrelation of the noise,
$C^{MM}_{\rm noise}$.  We calculate $C^{LL}_{\rm noise}$ in a similar
way by randomizing our catalogue of predicted shear values.

\begin{figure*}
\centerline{\epsfig{file=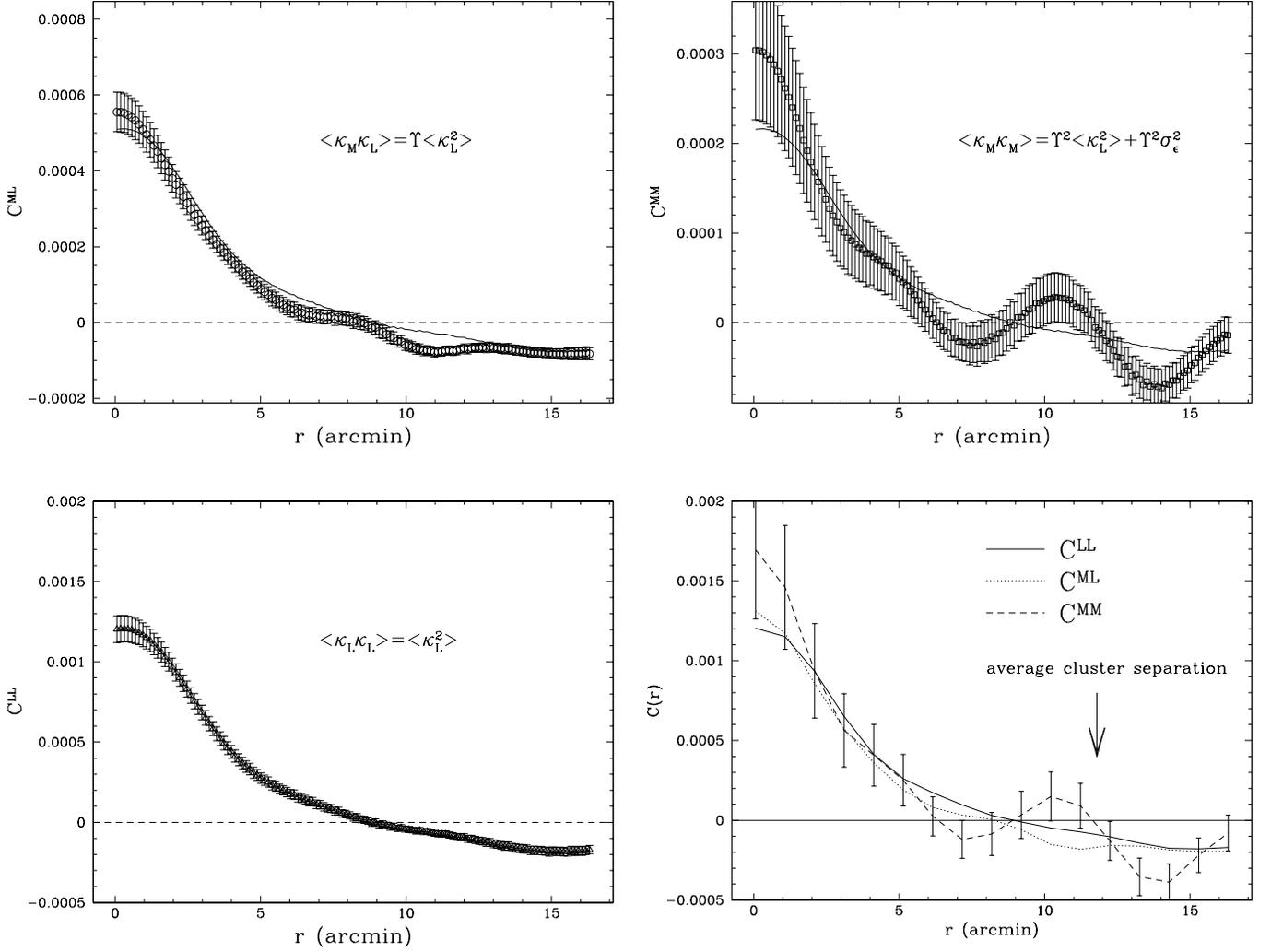,height=1.0\textwidth,angle=270}}
\caption{The radial dependence of the two-dimensional
cross-correlation of mass and light, $C^{ML}$ (top left) and the
auto-correlations $C^{MM}$ (top right) and $C^{LL}$ (bottom left).  In
each case, the solid curves show $C^{LL}$ renormalized according to
equations~(\ref{eqn:cml})--(\ref{eqn:cll}) and the best-fit values of
$\Upsilon=0.422$ and $\sigma_\epsilon^2=0$.  The assumption of a
linear bias model breaks down on the scales of the average cluster
separation, $r\sim10$ arcmin, reflecting the significantly different
$M/L$ ratios for the individual clusters.  Note that the points are
not independent. The bottom right panel shows the sampled points from
all three curves used in the $\chi^2$ analysis, renormalized to agree
with $C^{LL}$.}
\label{fig:xcprofplot}
\end{figure*}
 
Fig.~\ref{fig:xcprofplot} shows the azimuthally averaged $C^{LL}_{\rm
true}$, $C^{ML}_{\rm true}$, and $C^{MM}_{\rm true}$ curves, 
with error bars calculated from the noise reconstructions.  Note the
significance of the correlation signal at small scales in each case:
e.g. for $C^{ML}$, the zero-lag correlation between mass and light is
significant at the 10.6$\sigma$ level.  On larger scales, the mass
auto-correlation, $C^{MM}$, displays oscillations not present in the
other two curves.

\begin{figure}
\centerline{\epsfig{file=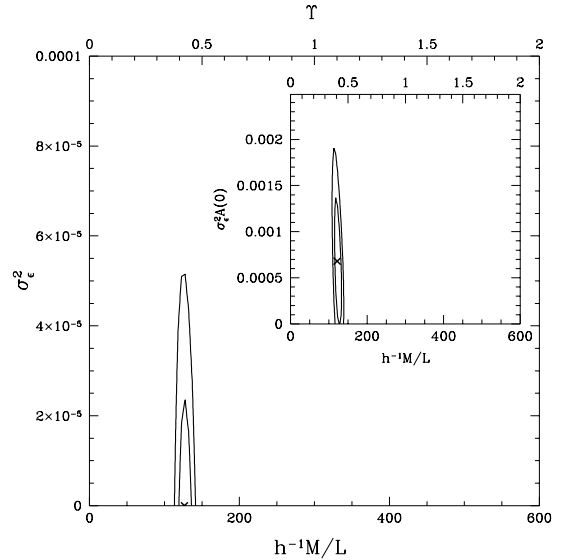,width=0.45\textwidth}}
\caption{Results of $\chi^2$ fit for parameters $\Upsilon$ and
$\sigma_\epsilon^2$ in equations~(\ref{eqn:cml})--(\ref{eqn:cll}).  The
best-fit parameters are $\Upsilon=0.442$ and $\sigma_\epsilon^2=0$,
corresponding to $M/L_B=126h$ and providing no evidence for a
stochastic bias component.  The contours show $\Delta\chi^2=2.3,6.17$
which enclose the 68\% and 95\% confidence limits for a two-parameter
fit.  Inset: the $\chi^2$ contours for a Gaussian model for the
stochastic component, $\sigma_\epsilon^2 \rightarrow
\sigma_\epsilon^2 A(r)$, where 
$A(r) = \exp[ \frac{-(r-r^\prime)^2}{2 r_s^2} ] /\sqrt{2\pi r_s^2}$,
and $r_s$ is the smoothing scale (1 arcmin).  The best-fit value shown
corresponds to the zero-lag value $\sigma_\epsilon^2A(0) = 5.97\times
10^{-4}$, with the resulting $M/L$ ratio unchanged.}
\label{fig:chisq}
\end{figure}

To recover values for $\Upsilon$ and $\sigma_\epsilon$, we proceed
with a $\chi^2$ test on equations~(\ref{eqn:cml})--(\ref{eqn:cll}).  The
points on the correlation functions are highly correlated, so we
sample the curves at 60 arcsec intervals (the Gaussian smoothing
width) to approximate independence.  The parameters producing the
minimum $\chi^2$ value are $\Upsilon=0.442$ and $\sigma_\epsilon^2=0$,
i.e. a $M/L_B=126h$ and no detection of a non-zero stochastic
component. The $\chi^2$ distribution is shown in Fig.~\ref{fig:chisq}.
For comparison, $C^{LL}$ is renormalized according to
equations~(\ref{eqn:cml})--(\ref{eqn:cll}) using the fitted values for
$\Upsilon$ and $\sigma_\epsilon$, and the resulting curves are shown
by the solid line in each of the first three panels of
Fig.~\ref{fig:xcprofplot}.  The lower right panel of
Fig.~\ref{fig:xcprofplot} shows the sampled points used in the
$\chi^2$ analysis, renormalized to agree with $C^{LL}$.

At zero-lag the agreement between the three curves is not particularly
good, and the amplitudes of the curves at $r=0$ imply $M/L\sim
150h$. Neither the $\chi^2$ nor the zero-lag value for the $M/L$ ratio
is in good agreement with the much higher $M/L$ ratios tabulated for
A901b and A902 in Table~\ref{tab:obs}.  Furthermore, the minimum value
of the $\chi^2$ function for the 17 sampled points is 34, indicating
that the linear bias model is not a good choice.  The model breaks
down to a greater degree on larger scales.  In particular, the
$\kappa_M$ autocorrelation function $C^{MM}$ shows a prominent
secondary bump at $r\sim11$ arcmin ($\sim$ 1 Mpc) that is missing from
$C^{MM}$ and $C^{LL}$.  This scale length roughly corresponds to the
average separation of the clusters and reflects the variation in
luminosity (and $M/L$ ratios) between the clusters.  While the peaks
are of similar size in the mass map $\kappa_M$ (producing the
secondary peak in $C^{MM}$ at $r\sim11$ arcmin), the peaks in the
mass-from-light map $\kappa_L$ vary in amplitude, most notably for the
relatively underluminous A901b.  Thus the assumption of a linear
biasing model with a single $M/L$ ratio describing the entire
supercluster fails on these scales.  Nevertheless,
Fig.~\ref{fig:xcprofplot} shows that the light from the early-type
galaxies and the surface mass density agree surprisingly well within
the errors.

In response to the poor agreement of the correlation function on small
scales, we consider an additional model for the stochastic component:
a $\delta$-function at zero-lag that is modified by a Gaussian
function representing the smoothing scale of the light and mass maps.
A non-zero measurement of a stochastic term will therefore likely be a
measure of the scatter in the $M/L$ ratio of the clusters themselves.
In this case, the autocorrelation of the stochastic component at
separation $r-r^\prime$ is
\begin{equation}
\lgl \epsilon(r) \epsilon(r^\prime)\rgl =
\sigma_\epsilon^2 \delta_k(r-r^\prime),
\end{equation}
and so 
\begin{equation}
\sigma_\epsilon^2 \rightarrow \sigma_\epsilon^2
\frac{\exp\left[{\frac{-(r-r^\prime)}{2r_s^2}}\right]}{\sqrt{2\pi r_s^2}},
\end{equation}
where $r_s = 1$ arcmin is the smoothing scale for our analysis.  Using
this model, the results of the $\chi^2$ are shown in the inner panel
of Fig.~\ref{fig:chisq}.  We find a non-zero value for the stochastic
component at zero-lag and an unchanged $M/L$ ratio.  However, the minimum
value of the $\chi^2$ function (31 for 17 data points) demonstrates
that the fit is only marginally improved, and so the detection is
tentative at best.  A more sophisticated model (with, for example, a
varying $M/L$ ratio) would be more likely to improve the fit.

\subsubsection{Cross-correlation of individual clusters}

\begin{figure}
\epsfig{file=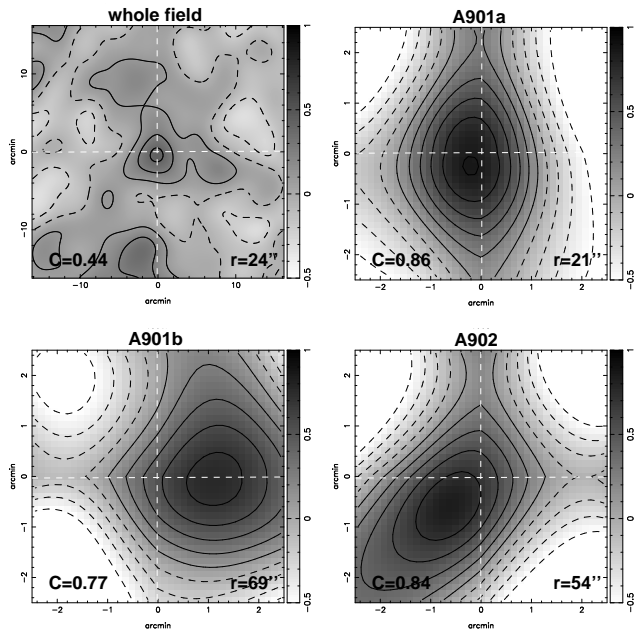,width=0.45\textwidth}
\caption{Two-dimensional cross-correlation of the $\kappa_L$ and
$\kappa_M$ for the whole field (top left) and for pixels extracted
from the boxes around each of the three clusters in
Fig.~\ref{fig:lightmasscont}.  The misalignment of light and mass in
A901b is especially evident, as is the elongation of the mass
distribution in A902 with respect to the light.  Values of $C$ and $r$
give the magnitude and offset of the peak of the correlation
function in each case.}
\label{fig:smallxcorr}
\end{figure}

To further explore the degree of correlation between light and mass on
the scales of the individual clusters, we perform cross-correlations
of $\kappa_L$ and $\kappa_M$ on pixels extracted from boxes around each
of the three clusters (shown in Fig.~\ref{fig:lightmasscont}).  As in
this case we are concerned not with the normalization of $\kappa_L$
but with the spatial alignment between the two distributions, we use a
cross-correlation estimator normalized by the variances of the
overlapping pixels at each image offset.  The resulting
two-dimensional cross-correlations are shown in
Fig.~\ref{fig:smallxcorr} for the entire field and for the three clusters.
The misalignment of light and mass in A901b is especially apparent,
while A902 also shows an elongated mass distribution with respect to
the light.

The correlation coefficient, $C$, can be written as
\begin{equation}
C= \frac{\lgl \kappa_M \kappa_L \rgl}{\sqrt{\lgl \kappa_M^2 \rgl \lgl \kappa_L^2 \rgl}} =
\frac{1}{\sqrt{1+\sigma_\epsilon^2/\lgl \kappa_L^2 \rgl}}.
\end{equation}
For Fig.~\ref{fig:xcprofplot}, the zero-lag values give $C=0.916$, or
$\sigma_\epsilon = 0.04 \lgl \kappa_L^2 \rgl^{1/2} \simeq 5.3\times10^{-4}$,
while for Fig.~\ref{fig:smallxcorr} we find $C=0.44$ for the entire
field.  This apparent inconsistency could be due to the fact that for
the cross-correlations of Fig.~\ref{fig:smallxcorr} the contribution
from to the noise was not removed 
(cf. eqs.~[\ref{eqn:clltrue}]--[\ref{eqn:cmltrue}]), reducing the
magnitude of the correlation.

\section{Discussion and conclusions}\label{sec:discussion}

One of the most striking aspects of the A901/902 supercluster is how
it reveals itself in many different guises according to the
observations at hand.  Were one to take the revised X-ray luminosities
of \cite{schindler00} at face value, it would appear that only one
moderate X-ray cluster is to be found in this region, at the location
of A901b.  When one examines the number density of color-selected
galaxies (Fig.~\ref{fig:lumndens}, left), however, the picture becomes more
complex.  One would conclude \cite[e.g.][]{aco89} that
there are in fact two clusters in the field (A901a and A902 in our
nomenclature).  A third overdensity near the location of A901b could
be considered, but it appears not nearly as prominent as the other
two.

Weighting the number density of early-type galaxies by luminosity
(Fig.~\ref{fig:lumndens}, right) changes the picture still further.  Now it
is A901a that leaps to prominence, dwarfing the other two clusters
with a compact collection of bright early-type galaxies.  However,
factoring the fraction of blue galaxies missed by the color-selection
could boost the total luminosity of A902 due to the relatively large
population of bright blue galaxies in its vicinity.

How does one make sense of these conflicting portraits of the
supercluster?  In this case, gravitational lensing provides a direct
link to the underlying mass distribution, and hence to theoretical
predictions of structure formation.  It also provides an additional
complication, as in the mass map (Fig.~\ref{fig:redfaintmass}) one now
sees three strongly detected mass peaks.  The relative strength of
each of the lenses (Fig.~\ref{fig:encmass}) is contrary to the number
density, light, or X-ray predictions, with A902 appearing the most
massive and A901a the least.  Furthermore, the mass distribution of
A901b is significantly misaligned with the early-type light
(Fig.\ref{fig:lightmasscont}).

We used a statistical cross-correlation to examine the relationship
between the early-type galaxy light distribution and the underlying
dark matter.  The light and mass are strongly correlated at the
10.6$\sigma$ level at zero lag.  Despite the fact that the clusters
exhibit a range of $M/L$ ratios (total mass to early-type light), we
find that the simple linear biasing model yields somewhat good
agreement between the cross- and auto-correlations of mass and light,
and do not find evidence for a stochastic component.  The best-fit
parameters imply that the mass is well traced by the light from the
early-type galaxies with $M/L_B=126h$, which is an underestimate
compared with the average $M/L\sim 200h$ computed locally for the
clusters themselves.  This could reflect that there is more light than
mass extended throughout the field, but is more likely due to the fact
that on large angular scales (on the order of the separation between
the clusters) and on small scales, the relation breaks down and the
linear biasing model itself is not an adequate description.

While the $M/L$ ratios of the clusters do not increase at large scales
(Fig.~\ref{fig:ml}), a single $M/L$ ratio is not an appropriate
description, despite their proximity to one another.  There are a
number of possible ways to reconcile these differences.  The first
option is to attribute it simply to a difference in the mix of galaxy
populations in the various clusters.  For example, we could modify the
observed $M/L$ ratios (which consider only the luminosity of the
color-selected early-type galaxies) by including the contribution of
the late-types galaxies.  This would require an estimate of spectral
or morphological type and a confirmation of supercluster membership
via redshift information, all of which will possible with the
forthcoming photometric redshifts soon to be available for this
system.  In this way, we could develop a more complete picture of the
galaxy populations for each cluster, and test if the mass-light
correlation for spirals is more extended than the strong correlation
between mass and ellipticals we detect here.  As we have seen in
\S\ref{sec:fb}, the galaxy populations also appear to vary from
cluster to cluster.  The potential large blue fraction of galaxies
around A902, if attributed to late-type luminous cluster members,
could boost the luminosity of the cluster and bring its $M/L$ ratio
more in line with that of A901a.  However, this would still leave the
anomalous case of A901b, with its high $M/L$ ratio and misaligned
distributions of mass and light.

A second possible explanation for the non-uniformity of the system
could be the effects of on-going mergers and interactions between the
clusters, which are in close proximity to one another (the average
separation being $\sim$1 Mpc).  We have seen in the mass map that a
filament of dark matter may connect A901a and A901b, and is mostly
unassociated with light from the early-type galaxies.  However,
Fig.~\ref{fig:galaxies}a shows that some bright galaxies do lie in
that region and were too blue to survive the early-type color
selection.  In this case, a spectroscopic survey of bright galaxies of
the field could look for signs of enhanced star formation in this
region.  Perhaps the extra luminosity of A901a relative to the other
two clusters, or the presence of bright blue galaxies along the
filament, is attributable to enhanced star formation triggered by
cluster infall or cluster-cluster interactions.  This would bias the
$M/L$ ratio of A901a low relative to the other two clusters.

Finally, some form of \textit{non-linear} biasing may be required to
relate the light and mass, with the $M/L$ ratio being some more
complicated function of the local density field.  We note that for our
supercluster, the most massive clusters have the higher $M/L$ ratios,
and the least massive cluster has the smallest $M/L$ ratio for all
aperture sizes, implying a mass-dependent scaling relation between
mass and light.  This highlights the importance of having a
comprehensive \textit{mass-selected} sample of galaxy clusters to test
the dependence of the $M/L$ and other properties on cluster mass.

\cite{kaiser98} presented the first weak lensing study of a
supercluster, in that case a system of three massive X-ray clusters at
$z=0.42$.  Our study complements this by examining a similar system,
but one that is less massive, at lower redshift, and more compact.
Our results differ in that we do not see nearly as strong agreement
between $C^{LL}$ and $C^{ML}$ as in their study (e.g. their
Fig.~20). We note, however, that our approach differs by removing the
contribution of noise from the correlation function (cf.
eqs.~[\ref{eqn:clltrue}]--[\ref{eqn:cmltrue}]).  They also do not
include the mass auto-correlation function, $C^{MM}$, in their
analysis, which in the case of the data presented here (see
Fig.~\ref{fig:xcprofplot}), displays significant oscillatory structure
that challenges the linear biasing model.  It remains to be seen if
their single $M/L$ ratio would simultaneously satisfy all three
correlation functions: $C^{LL}, C^{ML}$ \textit{and} $C^{MM}$.

One of the most striking implications of the \cite{kaiser98} results
was the low value of $\Omega_m$ determined under the assumption that
their measured $M/L = 280h$ is universal.  Assuming that early-types
contribute 20\% of the total luminosity density of the Universe, they
find $\Omega\simeq 0.04$ for a Universe with critical $(M/L)_{\rm
crit} = 1500h$.  This would imply an extremely low density Universe,
with no need to invoke large amounts of non-baryonic dark matter.  A
follow-up study by WLK attempted to further investigate these findings
with an identical weak lensing and cross-correlation analysis.  The
targets in this case were six blank fields, chosen to be free of any
known structures and thus potentially more representative of the
universal $M/L$ ratio than dense environments that may be biased or
display complicated environmental effects.  Following a similar
analysis, they found $M/L_B \simeq 300 \pm 75h$ for a flat
($\Omega_m=0.3, \Omega_\Lambda=0.7$) cosmology, and $M/L_B \simeq 400
\pm 100h$ for Einstein-de Sitter, consistent with the
\citeauthor{kaiser98} results.

Rather than estimating the fractional contribution of the early-types
to the luminosity density of the Universe, WKL directly integrated
over the type-dependent luminosity function from the 2dF redshift
survey \citep{folkes99}.  For $M/L_E = 300h$, this yields $\Omega
\simeq 0.1\pm 0.02$ ($\Omega \simeq 0.13 \pm 0.03$ for the WLK results
in an Einstein-de Sitter Universe).  These are higher than the \cite{kaiser98}
supercluster results and do require some component of
exotic matter to contribute to the total mass density.  However, we
note that were we to apply the same, more rigorous calculation to the
\citeauthor{kaiser98} $M/L=280h$, this would yield $\Omega_m=0.09$, more than
twice their original estimate.

In summary, both the \cite{kaiser98} and WKL studies find that
most of the mass in the Universe is associated with early-type
galaxies, and in the case of the \cite{kaiser98} supercluster the
mass is no more extended than the distribution of the early-types.
The low values of $\Omega_m$ obtained in this fashion assume that
little mass is associated with late-type galaxies (under the
assumption that an extended distribution of massive spiral galaxies is
not reflected in the concentrated distributions of mass), and so
assigned a negligible $M/L$ ratio to the late-types.  This assumption
is supported by the findings of \cite{bahcall95} who
find ellipticals to have four times the mass of spirals for the same
luminosity.  Our conclusions are roughly similar but with some
important differences.  Firstly, we do not see nearly the same
agreement between mass and light in our supercluster relative to the
\cite{kaiser98} study, both in the spatial distributions
(e.g. A901b, Fig.~\ref{fig:smallxcorr}) and in the $M/L$ ratios
(Fig.~\ref{fig:ml}).  Secondly, the global $M/L$ ratio (considering
the total mass and the early-type light) that we determine from the
$\chi^2$ fit to the cross- and auto-correlations functions is lower,
$M/L_B \simeq 130h$, which, following the same arguments as above,
would yield $\Omega_m = 0.04$.  However, this rests on the assumption
that the linear bias model is correct and that the global $M/L$ ratio
we measure is representative.  The failure of the correlation
functions to agree in shape and amplitude on small and large scales
shows that this assumption is not valid.

The clusters in the supercluster presented here are clearly not
isolated nor relaxed systems.  The effects of a pre- or post-major
merger state could be invoked to explain the discrepancies between the
alignment of mass and light.  In both distributions, evidence for
inter-cluster material is seen in the form of filamentary structures
between A901a and A901b, and some degree of non-linear biasing is likely.  

A wealth of additional information about this system is available in
the form of the photometric redshifts, which will be derived from the
remaining 15 filters in which this field was observed.  Aside from
allowing an independent measurement of the mass distribution by
measuring the effects of the gravitational magnification on the
luminosity function of the background galaxies, the redshifts will
allow for a more accurate characterisation of the foreground structure
than the relatively crude color-cuts employed here.  Furthermore,
rather than concentrate solely on the color-selected early-type
galaxies, a more comprehensive picture of the mass-light relation will
be constructed by separating structures in redshift space and by
spectral class.  This could include compensating for the effects of
foreground or background structures along the line of sight to
disentangle projection effects that could bias the two-dimensional
weak lensing mass measurements.  Additionally, one could perform
correlations between mass and spectral type to determine how much mass
is in fact associated with the late-type galaxies.  Finally, a
spectroscopic survey of the bright galaxies in this region could
provide the final piece of the puzzle, by providing dynamical
information and tracing star formation as a function of environment
(i.e. in the filament or infall regions).

\section*{Acknowledgments}
The authors would like to thank Nigel Hambly for assistance with the
astrometric issues and for providing the SuperCosmos data for this
field.  We also thank Lutz Wisotzki for obtaining the
spectrophotometric standards in the supercluster field.  We are
grateful to Alexandre Refregier, Omar Almaini, and David Bacon for
helpful discussion and comments.  MEG was supported by funds from the
Natural Sciences and Engineering Research Council of Canada and the
Canadian Cambridge Trust.  ANT was supported by a PPARC Advanced
Fellowship.  SD is a PDRA supported by PPARC.

\end{document}